# NARRATIVE BRIDGING

## a specification of a modelling method for game design

Katarina Borg Gyllenbäck

Dept. of Computer and Systems Sciences
Stockholm University

This thesis corresponds to 20 weeks of full-time work.

# Acknowledgements

I would like to thank:

- Magnus Boman: for always believing in me and for being a patient listener;
- the students: for their invaluable opinions and for lending their works to the study;
- Swedish Institute of Computer Science (SICS): for offering me a peaceful place to work;
- Peter Gärdenfors for enthusiastic and interesting conversation;
- Eric-Oluf Svee for invaluable, brilliant and skillful help;
- Marie Sjölinder: whose advice helped me prepare the study;
- Jussi Karlgren for his brilliance when helping me name the method;
- Annika Waern: for criticism that helped refine the final result;
- Helena Ruge for reading through the first drafts and providing feedback;
- Anette Sandegard who once drew me into the whole.

With love to family and friends.

**Do not show, involve**

# Abstract

Very little has been explored about the narrative as a process when constructing entertainment for interactive media. Simultaneously, the interest in narrative vehicles increases while certain occupations, seeing the narrative as a structure, obscure the examination of the process of selecting, arranging and rendering story material. To correct this deficiency, a method for a narrative bridging that encourages research and design while exploring narration as a process, is proposed with the aim to not diminish the properties of the interactive media. This method focuses on the initial phase where establishing and handling the information takes place and creates a foundation that precedes its systematization and computation. The aim is to give designers a comfortable design tool that firmly aids the design without interfering with creativity, and at the same time aids the construction of interplay between narration, spatiality and interactivity. The method aided the practise of a discipline that was established through a qualitative study conducted as part of a university course in rapid prototyping. The results demonstrated that the method aided time-constrained design processes, simultaneously detecting inconsistencies that would prevent the team from making improvements. The method gave the team a shared vocabulary and outlook, allowing them to progress without interfering with the creative flow. This enabled the team to reason about the process and easily advice design stakeholders. The study also provides directions for future developments within research of narrative processes in game design.

# Table of Contents

# 1 Introduction

## 1.1 Narration: an uncharted area within interactive media

In 1999, interaction designers contacted me in my capacity as a scriptwriter. They had problems with a computer-generated character in a conversation program (an agent) that was supposed to be used in a theatre performance [8]. During testing, it was discovered that the users did not find the agent engaging, an issue I was asked to correct. Nine years later I encountered researchers [9] who were developing another agent for educational purposes at the Royal Institute of Technology in Stockholm. Once again, the agent was not enjoyable to interact with and they asked for advice about how to create and manipulate the content to make it more attractive.

Although interactive media is technically a well-developed industry with many different platforms, in situations when digital media approximate entertainment, the industry and researchers face problems; there are no clear algorithms or manuals about how to trigger entertaining mechanisms or how to narratively iterate the contents to suit a desired outcome and platforms. A bridge to the narrative seems to be needed but the question is: How? David Bordwell [7] gave theories of narration a richer meaning as he criticized assumptions about narration and insufficiency within cinema. Bordwell found that most of the focus lay on techniques, photography and on isolated narrative devices at the expense of the entire work. Bordwell's criticism and reason for writing the book mirrors the current situation with how narrative theory and practise are treated within interactive media. Bordwell's solution to the reigning situation within film was to present narration as a process in contrast to culturally strong structures (the canonical story format), thus showing its interrelating systems.

Within game design, Chris Crawford [12] criticizes how designers use graphics to refer to the process of game design. He states that it is often only "eye candy" for merchandising purposes and that technical attributes should only be there to further the game experience (the game play). When it comes to narration within game design Henry Jenkins [20] states that if game designers are going to tell stories they should tell them well. Designers are very well schooled in computer science and graphic design, he says, but need to develop a basic vocabulary of narrative theory. Jenkins points out that the preoccupation with the rules and conventions of classical linear storytelling prevents a full understanding of the interplay between narrative elements. He suggests one should work with the term *spatiality* and argues that game designers should be seen less as storytellers and more as story architects focused on the interplay between narration, interactivity and spatiality. Therefore, the question is: Where does this pre-occupation and use of classical linear storytelling come from and how does it prevent an understanding for the interplay between narration, spatiality and interactivity?

The canonical story format, what Jenkins refers to as the classical linear storytelling [20], is a western cultural structure [7]. Its dramaturgy and established genres force creators to remain within a certain form, primarily because the previous experiences of audiences lead them to expect it. The reason why designers are occupied by these strong structures can be explained within remediation. Remediation is a concept and a method for looking at new and old media and how they influence each other, i.e. interactive media creators being inspired to do interactive television or videogames adapting the concept of a Hollywood film (Aristotle's plot of events towards a climax [7]) [6]. Within game studies it is a well-known phenomenon as one remediates activities between play and games in e.g. the comparison between digital technologies or toys [26]. But how structures influence practise in game design is an unexplored area. By looking at the canonical story format (exposition, complication and outcome) one can see how it affects game design that turns theatre and films acts into levels and how one lets a user shoulder the role of a hero to face bosses and tasks as in films. Within research into the practises of game design and narration, Vladimir Propp (1895-1970), a Russian literary scholar and structuralist [4, 7, 23, 28], has gained popularity for his deconstruction of folktales and transcription of all patterns and combinations, along with his mapping of characters as the hero and villains [4, 25]. Joseph Campell (1904-1987) was also a structuralist and influenced game design and research through his book, "The hero with a thousand faces" [10] where he described structures in steps and sequences of a hero's journey to reach a higher self [2, 3, 24]. While these are all phenomena based on the canonical story format and dramaturgy, there is nothing inherently wrong using strong structures or templates. But if these structures are applied to all games the risk is that one ends up with a one-sided market.

Marie-Laure Ryan's [29] advice is that one should not mimic film or literature and instead regard narratology as an unfinished project that should be expanded beyond the classical structures. Ryan's advice seems problematic if one likes to explore what lies behind the strong structures. Several books on storytelling and plotting for interactive media and games have been published in the last two years although most of them follow the schemata according to the Hollywood film concept, a sublimate form. One example is a book written by Marianne Krawczyk and Jeannie Novak [24]. Their advice on how to make a videogame is to watch films and read books to learn how authors create a story, make a premise, write a backstory, use the three-act structure and plot three levels to the game with plot twists, etc. All books have some good points to add to the discourse, but they also shut off the possibility to explore narration as a process. The process is how to cue, channel and manipulate pieces of information, whose main purpose is to make sense of a fictive world. These can be applicable in any media, depending on how one chooses to put them together. This is also what Bordwell suggests when offering narration, seen as a process, as a means of to getting around the problems of how strong structures influence the study and practise of narration [7].

Another problem with looking upon narration as a structure emanates from the fruitless discussions about narration being linear, thus preventing users from control. Researchers such as Jesper Juul have nourished the idea of narration as a linear phenomenon. In his 1999 thesis "A clash between game and narrative" [22], Juul's conclusion is that traditional story themes in books and the game-related phenomenon, as reaction time and skills in games, are not compatible. His comparison, when seen from a narrative perspective, will fail as he has compared two different media with different ways of structuring the narrative. Several theorists, when forming a theory for games, use narrative terms incorrectly when defining their hypothesis. One example is the use of the term "story" as something whose linearity while being told prevents users from control. It is correct that a story represents a set and fixed event that can be retold and referred to. The problem occurs when using the term "story" to allude to the process of constructing a story. The practise is a construction of causal, spatial and temporal links involving the receivers work "reading" the information. These procedures can be explained by other terms – the *syuzhet* (plotting) and *fabula* ("reading" of the information) [7] and are adapted to the desired outcome and media. If using the right terms, game theorists like Espen Aarseth [1] would also have a vocabulary when criticising film and literature theorists for their misapplication of narration on games instead of calling their efforts an incompetent outnumbering of the ones that understand what games are.

Henry Jenkins suggests solving the problem, when new and old media collide, through collaboration and education [21]. If one looks at higher education within game development, current theory even biases the instruction as the programmes have hardly any courses in narrative construction, their main focus being on graphics and programming. Students are given textbooks that are either influenced by strong structures or present the narrative as an appendix to all other systems within game design. The question should instead be why narrative aspects have been omitted and criticised by a majority of experts within the design of videogames and yet still influence a majority of the education? The academic world has spent over ten years attempting to define the digital game and it has resulted in an interdisciplinary disagreement about whether digital games should be considered as rule-based or narrative [30]. Currently one can sense that there is a kind of agreement that narrative and game rules do support and overlap each other. The focus instead seems to have shifted to the question of which one is superior in the design of digital games by debating "games as stories or stories as games" [30], p. 379, which leads back to square one as the term "story" is obscured.

Crawford [12] thinks it is good to have designers experienced within both art and technology to influence future game design. But the question is how, for example, a scriptwriter shall approach the technology-based area and how the market looks upon the artist's skills? Juul [23] has chosen to not use the term narration in his PhD thesis. Instead he argues that, when handling the narrative components in videogames, the term *fiction,* taken from philosophical theory, should be used. Fiction means something that is "made up and not real" and by using this term, he excludes the mechanism behind what actually "makes" the fiction–the narrative construction–and says that games cannot handle complex themes as love, social conflicts, etc. How can we be certain about that if narration is not treated with correct terms or explored as a process? By using the term "fiction" Juul also spreads a notion about the narrative arising from an artistic condition and being performed by persons with great imaginations. There is a risk that this notion will be confirmed if narrative theory is not explored or seen for what it is and a considered artistic profession treated as an arbitrary practise. Within any art form, including the design of interactive media, there exist forms of pre-knowledge and intuitive practise. Intuition and pre-knowledge are not bad, but when handling large complex systems, not being aware or not knowing how the narrative mechanism works affects the work and one runs a risk of moving the idea towards manuscripts, systematization, graphics and computation too quickly. These problems may be hard to adjust or detect, which again could lead to the question: "Why didn't the user find it fun?".

## 1.2 Narration as a motivating vehicle

The most common form of entertainment within interactive media is the digital game. A multifaceted phenomenon to define, just looking at all the different definitions of a game says something about its complexity. Salen and Zimmerman are described as formalists by Frans Mäyrä [26] as they define games as system-oriented. The systems represent a player involved by rules of conflicts that, to create a meaningful play, has several outcomes. Salen and Zimmerman [30] in turn gathered eight different definitions made by influential experts within game research and design and ranked fifteen of their characterizations to see how many corresponded to each other. Four definitions where shared by most of them as "games are rules that limit the players", "goal-oriented/outcome oriented", "involve decision-making" and are an "activity, process, or event" [30], p.79. These are theoretical descriptions of digital games but Salen and Zimmerman also describe the practise for how to create meaningful play as follows [30], p. 34:

> "Meaningful play occurs when the relationships between actions and outcomes in a game are both discernable and integrated into the larger context of the game. Creating meaningful play is the goal for a successful game design".

So what is it to create meaningful play? How can the digital game, as a representation for interactive media, cooperate with the narrative? Or rather, how shall one enable narrative bridging? When it comes to the construction of the agent that was not fun to interact with, the term "play" might not be the first thing one thinks of. But if we agree it is so, the question might not be how to make the agent fun, but rather how to convey information that motivates and engages a receiver to take action. An explanation of this can be found in cognitive theory, which deals with the mechanisms behind the creation of a meaning.

Peter Gärdenfors [16] outlines the constant human search for meaning by looking for patterns to understand the world. In philosophy, meaning has been defined, but no one seems to have cared about how it arises, according to Gärdenfors. He talks about the evolutionary cognitive vehicles behind human beings searching for meaning as part of our goal to survive. Storytelling was developed from the moment we got a language, as we needed to cooperate and fulfil goals. Humans also became good at simulating situations to detect danger or weighing possibilities by evaluating causes and probable outcomes. Different goals emerged to choose between, through the ability to simulate situations. This ability to make choices demands a rich inner world. The advanced planning even includes the ability to take decisions about several events connected to each other and the inner simulations continue to find new consequences, according to Gärdenfors. The human search for meaning and the ability to create new consequences demands empathy to enable an idea about someone else's consciousness: we can tell if someone is hurt and relate to our own inner representation of it. We can also tell what other people focus on by seeing where they fix their eyes and we can imagine someone else's thoughts (not necessarily correct but yet we do). This empathic ability is important and enables us, through the ability to separate one's own feelings from others, to understand someone else. If someone does something we generally believe that it was intentional and that was a cause behind the action. We easily imagine the aim of the action even if it does not correspond to the other person's intention, [16]. What we do not understand we fill in, in order to make sense of a context, which again directs our thinking towards a pattern of meaning.

Humans are goal-oriented, seeking meaning through causes and patterns and seeing consequences and making choices by understanding a context. These mechanisms are so strong that even when a receiver is not given any information, this need to create meaning will still take place via finding patterns [16]. These cognitive vehicles are shared in both game design and narrative practise by giving the receiver a context with a goal, generating causes and effects, creating patterns, and opening up choices within a context. Both narrative practise and game design share causality, cause-and-effect, as well as constructing interrelationships between objects and events. All of these bring logic and rules and consequences to a fictive world for a receiver to take action upon. These principles are reminiscent of a scriptwriter's practise of lining up information to suit a certain premise towards both a desired outcome and media, and it is through this perspective one can access the narrative process.

So narratively balancing a game is about creating meaning so that the receiver can cognitively create patterns and strategies for choices. Then the system will offer outputs that confirm the system's intentions and in that way encourage activities to take place. And as a reply to Salen's and Zimmerman's definition to make the narrative process more clear in creating meaningful play, I would offer following definition:

> Meaningful play occurs when causes and effects are elaborated by their creator towards a set premise and treated for the media it concerns, as well as when the effect of this supports the set goals for the receiver.

This is also the vehicle behind the method and the narrative bridging towards the interactive media that will be proposed in this work and described in depth through the second chapter. It is a vehicle that mimics and simulates cognitive patterns through causality, setting goals towards a desired outcome. This is to enable the receiver to see consequences from their own action in an environment and from that making decisions.

## 1.3 Purpose and goal

The purpose is to propose a specification for a method that remedies the deficiencies of the situation described above, hereby named narrative bridging. The function of the narrative bridging method is to aid the viewing, cuing, channelling and handling of contents before it is systematized, scripted and computed. The hope is to offer a comfortable method that comprises several systems, and aids in their control. The method aims to support the intuitive and creative processes and to constrain either the idea or the interactive media, while also balancing the creation of meaning for a receiver to cognitively create patterns and strategies for choices. Finally, the hope is to visualize the strength and workability of the narrative in the fields of education and research, as well as to provide the game design business with a tool to explore new expressions.

## 1.4 Methodology

This research builds upon David Bordwell's theorizations about the narrative as a process of selecting, arranging, and rendering the story material [7]. His work explains the process of constructing narrative information by presenting the systems and elements behind it and why narrative information is media independent (i.e. adaptable to various media). His research offers a cognitive, constructivist perspective, which aids the possibility of a user-centered view (see Mimetic theory, section 2.3.1). Bordwell's idea for narration is based on Plato's definition of narrative having someone speaking as if it were someone else, thus laying the groundwork for the diegetic theory of narration. In 1953, Etienne Souriau resurrected the term *diegesis* and it is now used to describe the fictional world of the story, e.g., its laws, rules and relations and how it is told. Early 20[th] century Russian formalists, in particular Viktor Shklovsky, Yuri Tynianov and Boris Eichenbaum, formulated criticism towards traditional narrative theories, something which was fundamental to contemporary narrative theory and Bordwell. They developed concepts as *syuzhet*, *fabula*, *motivation*, *retardation* and *parallelism*. Bordwell's definitions "fabula", "syuzhet", "premise", "cause-and-effect", "goal" and finally "the diegetic world" will be used to explain the method [7].

The narrative definitions and the formalistic way of approaching to a creation will be transferred and applied to interactive media, in particular the digital games that this work is directed towards. The theory of games and its framework for defining digital games are based mainly on Salen and Zimmerman's work specifying design, play, and culture with the purpose to create meaningful play [30]. Salen and Zimmerman's approach to create meaning in games is combined with cognitive theory, as Gärdenfors [16] carries it, explaining the mechanism behind human search for meaning. From there he explains how to create motivating patterns for the receiver, thereby enabling strategies for choices. This is the basis for explaining how narrative theory and game design can share the same vehicles. The vehicles are formed when mimicking cognitive mechanisms, creating an enhanced experience for a receiver by way of its causality, creating interrelating causes-and-effects, or creating and developing logic and rules that motivate an interaction. The narrative and cognitive features can be seen as elements of a bigger system that Tracy Fullerton [14] forward as an important way of looking at game design and whose view can also be seen throughout this work as a formalistic way of looking at the phenomenon which Fullerton shares with both Bordwell, Salen and Zimmerman.

A qualitative research project was conducted within a course that was part of a study programme in game design. The method was tried out during one of the course's sections, on the theme "Narrative". The test was held during a two weak period where the students were offered different methods to try out rapid prototyping of games (e.g. MDA [18]). The research project consisted of practise, interviews and questionnaires to establish whether the method for a narrative bridging worked and, if so, how it might be improved upon.

The study was carried out:
- via questionnaires, illuminating how narrative background and pre-knowledge worked, in addition to how it affected design practise and idea.
- in two iterations, where one was produced without guidance and another that was produced after a lecture where the method was provided. This was to generate material that would be used for evaluating whether the process was aided by the method.
- through interviews, a direct response and opinions about the design process and the use of the method.
- via seminar and lecture, paying particularly close attention to the students questions and opinions.

The first week, the students were given a task to create a game without any guidance and told to work consciously with the interplay between narration, game play, and game mechanics (hereafter named *style and mechanics*, see explanations in sections 2.4.3 and 3.8.3). In the second week, there was a presentation of the work, a lecture concerning narrative, and a presentation of the method. The students iterated their game ideas using the method and presented the new result a week later. The students were informed that the method was not perfected was offered to the research project as an instructional aid. The results were analysed by comparing before and after the method's introduction, using a questionnaire as well as interviews where students could offer their opinions about the method.

## 1.5 Delimitation

*What kind of games and media does the method support?*

The specification of the method for game design aims to support development of digital games with a narrative. Here the narrative means any game that has a narrative element that creates a context that shapes an interpretation of a fictional world. This mean that the method is not designed for use on abstract, strategic and graphical games like Tetris or chess, even if the method implies that it supports any digital game with a cognitively narrated element which could conceivably include "a king" or "a queen".

The study also touches upon the narrative as a representation, the semiotic and hermeneutic theory that permeates the narrative as a process for constructing logic, rules, and causality for a cluster of systems, which, along with their various elements, create motives for interaction. The semiotic and linguistic aspect of the construction is highly relevant for this study but those junctions of signs and nodes that create dynamics and tensions to an experience require a study of their own. One example is that the study cannot present how signs representing a genre trigger certain dynamics to create an emotion when making a thriller, etc.

In the previous sections, it was suggested that the method suited all interactive media with a narrative, but that cannot be proven within the thesis as only digital games were chosen as the example for testing of the method.

*What phase in the design process does the method support?*

This work focused on the narrative process for constructing a digital game beginning during the initial phase where the idea is born and its first representations are set. The conceptualisation and development of the ideas which are eventually systematized and computed are tiny parts of the entire process of making a digital game but is perhaps the most critical, for it is at this stage that all the components are set that establish the rest of the product. Additionally, the present work will not handle game development for tasks such as the design of health-bars, calibrating weapons, procedural characters, AI, levels, quests modelling, GUI, dialogues, etc. This thesis instead treats the representations that precede the decisions for how to model health-bars, etc., and ends where formats such as flow-charts, manuscripts, outlines, pitches, and UML have to be transformed into actual computing.

*Who can use the method?*

The goal with the method was to construct a workable tool for anyone involved with game design business and research and who is interested in handling and controlling the motivating and cognitive vehicles that the narrative offers. This work should be seen more as a possible way of teaching narrative and the method within game design programs. This fact does not exclude the possibility that the method would work in a real production environment but only that such usage would require further study.

This study, when it comes to the presented ideas and how to carry out the task and respond to the method, has delimitations. Perhaps the most important fact to take into consideration was the homogenous sampling in the study. There was only one woman in the study group and the individuals were between 22-29 years old. The ideas might have looked different if the subjects were exclusively women between 35 and 45, but this cannot be known. The homogenous group means that the viability of the method or its limitations and how it can be improved cannot fully be depicted.

*Theoretical gaps*

The construction of the method has some theoretical gaps when narrative bridging is defined, as when the transfer and alignment between the narrative and the media-specific attributes for the game and interactive media are presented by, e.g., transferring "space" to be represented by the term "world" (see section 2.3.1). In this example, it should be stressed that the transfers are made to illustrate the method and are not there to make theoretical claims.

## 1.6 Disposition

### *Chapter 2*
The extended background explains the systems and the elements behind the method. Piece by piece the terminology and theories behind the method will be presented to finally create a whole: the method known as narrative bridging. First the narration, spatiality and the interactivity systems will be presented. The terms *goals*, *premise*, *causes-and-effects*, the *diegetic world* and the *fabula* will also be introduced. Secondly the interrelation between the elements creating the interplay and causality will be explained and the narrative systems such as the *syuzhet* and *style*. These systems also explain why technically driven games, as a design goal, do not work if one likes to create a motivating experience, as well as how the narrative systems can collaborate with the game play and the game mechanics. Finally, a work specification will be presented for how to carry out the practise using the method.

### *Chapter 3*
This chapter will present how narrative bridging was initially tested. It will discuss about the challenges, processes, as well as the empirical data collection. Finally, the data from six groups that volunteered game ideas they generated within a course on game design and rapid prototyping will be presented.

## Chapter 4
Based on the empirical data collected, an evaluation of the results is presented where the good points, problems and improvements of the method are shown.

## Chapter 5

In this chapter, the results will be evaluated to see what needs to be contributed or withdrawn from the specification of the method to get a workable method. This chapter also presents additional results and advices for future progress within research of narrative processes in game design.

## Chapter 6
The discussion highlights further developments and possible procedures that can be used within design of interactive media to improve the usage of the narrative.

## Comments about expressions

Finally there are expressions in the thesis that are carefully chosen but need to be pointed out due to the existing ideas about narration. One expression is the definition "receiver" which is meant to be the representation of the person(s) taking part of the media and do not refer to a passive activity. At some places the receivers' activity will be expressed as the "receiver reads the information" and does not refer to text-based media. The "reading" refers to the activity of taking part of the informations, seen from a cognitive perspective.

# 2 Towards a method for narrative bridging

To enable a study of narration and its motivating vehicles one can start by looking at narration from several perspectives and then separate its functions. Bordwell [7] did just this, dividing narration into three perspectives. First is the *structural perspective*, which deals with canonical story formats, genres, and how the narrative is structured by different media. The second perspective is the *representation* of narration, which deals with the how to portray the events and their meanings in a broader sense. This touches also upon the semiotics and language, which deals with the signs used by people to designate objects or ideas that are significant to them. Those are then interpreted, resulting in a context that shapes an interpretation [7, 30]. Finally there is the *narrative process,* which deals with the construction of causal, spatial and temporal links involving the receivers work in "reading" the information. This is a set of interrelating systems that create logic, rules, and relations between the elements that generate an overall motivation for a receiver to take part in an experience. The construction is media independent and the "bricks of information" are arranged depending on both the target media and desired outcome. The formation of the current method is based upon this last perspective, aiming towards interactive media which has a narrative.

In short, the narrative bridging method is meant to support creation, organisation, control and generation of information for the interactive media. These are defined as digital games, interactive installations, narrative characters (agents), pervasive games, online games, etc. For the purpose of this thesis, the digital game is chosen to represent interactive media. The method will be presented throughout this chapter, starting with defining the digital game, which is the end result of the method. The systems behind the method will be presented as well as the narrative vehicles that make the method "run". Finally a graphical overview of the method can be seen and alongside a tripartite, prototypical specification for how to process the method. Next, the pieces, forming the specification of a method for game design of digital games with a narrative, will be presented (the final result for the building can be seen in section 2.5, Figure 8). In short the building of the method and its motor can be described as follows:

Narrative bridging works on the supposition that the interactive medium has a *user*, that there is *hardware* and *software*. The software is digital and created via programming. Programming provides order to conditions and choices by organising them through sequences, selections and iterations. Simply expressed *interactivity* can be seen as an action of input and output between user and the digital artefact and to have a variety of choices or walkabouts that demands *space*. The elements user, interaction and space will be expressed here as *character*, *world* and *action*, for the purpose of modelling interactive media that has a narrative. These elements make a base for causality creating interplay between *narration*, *interactivity,* and *spatiality* provided that there is a *premise* and a *goal*. The vehicles behind causality and the causes-and-effects, are generated with the help of the syuzhet principles that connect the elements between the systems. The systems are based upon the diegetic world that represents the logic, relations, and rules, interrelating with the spatiality and interactivity. The causality between the elements and the systems generates conditions and consequences for the player's experience, activity and game play. This finally makes a base for further systematisation, scripting and computing.

## 2.1 A method for design of digital games

The reason for naming the method "Narrative bridging" is that the method aims to support a design process for interactive media with a narrative. The method could also have been related to a "formal abstract design tool" (FADT) an expression proposed by Doug Church, meaning a tool that is "empty" thus helping designers build a digital game [11]. This tool, he suggests, should involve the player by having a design accumulate the goals, understand the world, make a plan, and act upon it. The tool is referred to as the *intention* to make a plan in order to understand the game play options. Church's message is that the solidity of a game makes the player feel more connected, and that the player should always be able to understand through their own actions why things went wrong. The tool is referred to as *perceivable consequence* to create a clear reaction from the game world upon the player's action, as seen in role-playing games, plot or character development. The best use of this tool was when the consequences were attached to intentional actions. As a third tool Church referred to the *story* as the narrative thread, design driven or player driven, that binds events together and makes the player move forward. It should be mentioned that Church also ends up in a discussion about narrative preventing the player from control of its environment and dissented from be pushed forward. The problem with Church's tool and other models is that they are hard to apply in practise and it is more of a design goal. Stefan Grunvogel [15] considers the drawbacks of formal models being that they force designers into a standardized workflow that they need to learn which is not naturally integrated in language. His holds the view that formalised methods should be more seen as theoretical frameworks. That is also the reason for why it is hard to apply Church's "tool" on this work, as his "tool" is more of a criteria instead of offering a strategy to obtain results.

To create a strategy it is important to know what the method aims for, as the narration is media independent and adjustable. Adams and Rollings [2] once defined the operational rules of a game world as *narrative* (without this the player will not be interested), *interactivity* (the way the player plays the game) and the *core mechanics* (the way the game works) [2]. The problem is that Adams' and Rollings' definitions do not allow for the narrative process to align causality between systems and elements as they treat narration as something told and fixed by saying: "Narrative is the noninteractive, presentational part of the story" [2] p.10. Tracy Fullerton [14] avoids conceptualizing the game world as Adams and Rollings do. To allow for creative game design she proposes to work with the physiognomy of elements in the game as a player's focus, setting of a premise, having an outcome, containing rules, conflicts, boundaries and elements of drama for the emotional experience. These elements together create a whole – a system – according to Fullerton. This is also an approach shared by Salen and Zimmerman that rather look at the form and what it contains [30]. These views also enable a connection to Bordwell's formalistic way of looking at narration as systems that needs to be processed [7].

The construction of the method is based on the narrative where characterisation is media independent and therefore needs some representative cues to aim the construction towards. The cues representing the media are here defined as the *narration*, *game play* and *game mechanics*. The *narrative* process is an action to outsource and align "bricks of information" a term which here means any object with an attribute that within semiotic studies, can be seen as a sign which people use to designate objects or ideas. These objects mean something to someone, are interpreted, and thus create a context that shapes an interpretation [30]. These elements are integrated into a bigger system that creates a whole and supported by the narrative systems as the syuzhet and style (see section 2.3.2).

Fullerton describes that narrative as "dramatic elements" which give a context to *game play* and integrate the elements of the system to create a meaningful experience [14]. The *game play* is a definition used within game design and there is no universal description for it, according to Adams and Rollings [2]. Richard Bartle regards game play as being what the environment offers the player [3]. Adams and Rollings say it is the challenges creating an action that makes the game play [2]. Salen and Zimmerman [30] see game play as a formalised interaction that occurs when the player is experiencing the digital game's rules and system and Fullerton [14] sees the game play as the core of the functions in a game.

Juul [23] regards game play as being what makes it fun to play. This work will be referring to game play similar to Fullerton's view of how systems and elements interact. That is as the interactivity that is formed by what the space offers the player via its narrative, and which creates the overall experience of the digital game.

The third element of the digital game is the element that supports the overall narrative and game play within the game context. It is here influenced by the expression *game mechanics*. The Game mechanics is also not a solidly defined term within game design even if it is frequently used when talking about game design. Fullerton [14] describes the phenomenon in terms of "core game play mechanism", or "core mechanic" and refers to the repetitive machinery creating and balancing negative or positive loops that the players partake in. Another definition used for the game mechanics is the expression *game engines* [19]. This is the mechanism marshalling all assets of a game, such as the AI, camera, platforms, graphics, lightning, texts, GUI, and features as the player's perspective, etc. In this work the game mechanics or the technical implementation of the contents will be given a new expression to clarify the purpose of integrating the narrative process and its systems. The intention is to demonstrate how the narrative objects and attributes affect and set rules for logic in the mechanics in the overall experience and these will be called *style and mechanics*. The definition will be described in depth later on in this chapter (see section 2.4.3).

*Narration*, *game play,* and *style and mechanics*, are considered to be the essential elements representing the framing of the whole digital game that the method is supporting. It is a way of directing the narrative process towards an end product due to the media's unique features. There is an internal order that has to be taken in to consideration when building the elements and these systems which will now be described in more detail in the following sections.

## 2.2 Defining the systems behind the method

Salen and Zimmerman define a system as *sharing elements that interrelate and effect each other to form a complex whole* [30]. Depending on how the systems are framed the elements differ. The elements are all objects that create a part, variable and relations in a context that surrounds the system. Salen and Zimmerman identify four system components: *objects*, *attribute*, *internal relationships* and the *environment*. If one looks at the interactive media defined by the elements *user*, *interaction* and the *space* as setting the framework for the systems *narration*, *game play,* and *style and mechanics* the internal relationship between these is now going to be defined as the systems creating the interplay between narration, spatiality and interactivity that can be set to suit the digital game: this is the narrative bridging method at its most essential.

### 2.2.1 Narration as a system

Within the theory formation of game design two kinds of *narrative* systems are defined and related to the interactivity, these being the *embedded-* and *emergent narrative*. Embedded narrative means that the user unlocks the contents bit-by-bit and is authored by the designer to frame the interaction e.g. when using of strong cultural structures as the canonical story format. The emergent narrative is a system which opens up several outcomes and can be played in different ways as when the player has an avatar in "The Sims" and takes care of it in a various ways. Games can use both emergent and embedded systems [30]. Many assert the emergent narrative to be the best system as it offers choices and represents the ultimate user-based design by letting the user e.g. take control over an avatar as in *The Sims* or when players in an online game forward the narrative through social interaction and role playing [27]. The only problem with the definitions of these systems is that when a genre is applied, it implies a certain pattern. The crime genre, for example, requires the content to be developed in a certain way, based on human pre-knowledge. This complicates the descriptions of systems being "open" and "closed" and could require a more detailed redefinition. Therefore these systems are only used as a general conceptualization of the games, and do not describe the foundation of the method even if supporting them both.

So if one looks at the narrative as a system the question is, what defines the parts and elements? Being goal-driven, the minimum required to motivate a human being to "read" information is a *goal* [7, 16]. The goal is also a fundamental element of the digital game and makes players attached to the outcome [12, 13, 14, 23, 30]. Seen from a cognitive perspective the goal- driven human reads all action as intentional [16]. This "reading" begins the process within the person to make sense of an action and construct a pattern, to understand the logic and rules to imagine the outcome and finally to make relevant decisions and choices from the conclusion [16]. These are the mechanisms that have to be kept in mind when constructing rules for a game and it is also these mechanisms the narrative practise tries to reconstruct, arrange and select whilst creating logic and rules for the fictive world (see Spatiality as a system, section 2.2.2).

When constructing the goal, the *conflict* is regarded as playing an important role in creating obstacles for the user [12, 13, 14, 30]. But as conflict can come in so many varieties, these occasion-based movements will be expressed here as a *"cause- and effect"*-construction. The minimum needed, to carry out a narrative construction, is a *cause-and-effect* and to create this causality, a *premise* is needed [7, 14]. Simply expressed, if wanting to construct a goal for the player one needs a premise to implement within the narrative system and its elements. When the goal and the premise are set the modelling can begin to create interplay between the elements by plotting the causes-and-effects. The mechanism behind the plotting will be explained by the three syuzhet principles in the section that treats the interplay between the systems (see section 2.3).

Thus, the narrative will be referred to as a system, describing how designated objects and their relational logic refer to each other. It will be explained how these elements construct a whole that encompasses the interplay between narration, spatiality and interactivity.

## 2.2.2 Spatiality as a system

Decision-making and having several choices within the space is a strong component within the digital games [2, 3, 12, 13, 14, 23, 30]. Salen and Zimmerman call this the "space of possibilities" and according to them, this can be seen as the actual *space* where the choices take place [30]. In the construction of the method space is where the elements of the narrative system (goal, premise, causes-and-effects) are presented and constrained and it is here that the player encounters the narrative information. Within narrative theory this space is referred to as the *diegetic world* [7]. Within the diegetic theory of narration the diegesis means the fictional world of the story that informs the player about logic, rules, laws and relations within the storyworld. When constructing the information for the diegetic world the practise is to define "how" and "where" to expose the information in the space and how, when, and where the "activity" occurs. The film "The Matrix" [33] serves as a good example of how spatiality and interactivity in a diegetic world can be narrated. In this film, a transport system has been constructed by using a telephone controlled by a person at a computer, thus enabling travel between two worlds (the real world and "the matrix"). Questions like *how* it works, *who* runs it and *why* it has to be used, provide information about the spatiality and interactivity. It also creates logic and rules for the diegetic world.

The compilation of these questions develops motives for a player to take action towards a goal.

The practise of how to forward and transport the narrative information to create the diegetic world is where designers easily end up showing the information as a background text or "cut scenes" instead of processing the narration through interplay between narrative, space and interaction (see Interactivity as a system, section 2.2.3). Within the movie genre, as in novel writing and narrative journalism, writers are advised: "Show, don't tell". Using the same idea in digital games design, the advice could be: Don't show, involve, meaning: let the player participate in the world instead of reading about it.

### 2.2.3 Interactivity as a system

Interactivity is the most apparent system within design of digital games. What the interactivity is varies within the definitions but has been seen as everything from what one can do, see, and hear in the game [2, 3, 12, 13]. Marie-Laure Ryan [29] suggests that interactivity can be seen as both an *internal* and/or *external activity* defining both the internal mode where the user controls and navigates within the software. Ryan's definition help explain the construction of those methods that focus on internal activity, how the inputs (design) and outputs (the interpretation of the design) within the software effect the receiver and how this then moves out to the final technical devices and the external activity.

From a narrative perspective the information to encouraging interactivity is an action delivering narrative objects and attributes into the diegetic world to create space for action. It allows for both the emergent and embedded narrative to be created and the reason for that can be seen in the syuzhet system, (see section 2.3.3 and 2.3.3) and that the plotting (syuzhet) adjusts to the media. It is a matter of deciding when, where or how the information shall be retrieved.

According to cognitive theory about human's search for meaning when not given information, or given abstract information as in more artistic driven events, the human will create meaning anyway [16]. If one wants the player to do certain things this "human-meaning-making-mechanism" has to be taken into consideration. The player's action and the outcome of it come about as the player interacts with the system. The goal designing interaction is to think about the meaningful choices [30]. These can be seen on a micro and macro level, according to Salen and Zimmerman, where the micro choices represent the small moment-to-moment experiences that are chained into larger macro structures of choices. From a narrative perspective Salen's and Zimmerman's description of the meaning of choices recalls the fabula which shows how small bits of information are chained and read by the receiver to finally end up at a whole that fits the overall goal for the receiver [7]. When it comes to digital games and creating interactivity the narrative system has to be set in a way so the player can "read" information, make conclusions, choose and decide what to do. The construction should reply to the player from the agreement of the premise and goal.

## 2.3 Align the systems for interplay

To substitute the systems of narration, interactivity, and spatiality, which would enable narrative bridging, the designer has to be able to imagine the reading of the information – the fabula – to form a coherent system where causes-and-effects can be constructed. Having experience from how to cue, channel and manipulate information helps the practise but is a good practise to keep the premise in mind. The premise is the initial seed that begins the process and works to guide the contents so the complexity of the interrelated systems does not become cumbersome. The aim when using the method is to become guide for the idea, whose interrelating mechanisms and causalities of the systems, will be presented in depth in the following sections to finally form a whole.

Organising the elements in the system is an act of balancing the logic, rules and relations that create the core elements in the diegetic world. Within digital games rules and relations create choices and outcomes. The construction of the diegetic world aims towards the same – to enable those choices. The organisation and balancing of these elements must make sense in total and be expressed so the player can experience the entire system provided that the narrative is processed for the media specific system and has a goal and a premise.

What lies behind the narrative bridging and its "vehicles" are four narrative terms which create motivational engines that enable a construction of information that makes a receiver to cognitively process the information (see section 1.2 Narration as a motivating vehicle). These "vehicles" are the *syuzhet*, *fabula*, *style* and the *diegetic world* (the diegesis*)* and are terms explaining the narrative building and manipulation of cognitive mechanism for providing the human to construct internal patterns [7]. But to make the systems of the syuzhet and the style work the diegetic world needs to be formulated.

### 2.3.1 Character, world and action

As said in the beginning of this chapter, the systems of the interactive media are here defined as having a *user*, *interaction* and *space*. These systems are represented narratively by the story components, which can also be called elements of information – *character*, *world* and *action* (see section 1.5, Delimitations). These also qualify the formulation of a diegetic world. The separation of the story components *character*, *action and the world,* which taken together represent the user, interaction, and space, was a strategy that turned out to be very helpful when constructing the agent (see section 1.1) [8]. As there were no avatars or any other representational images for the users that were simply visiting, the separation of the story components helped to see what elements in the system needed to be designed and controlled and what could not be controlled (e.g. the visitor).

*Character*
The character element is the user's/-s internally narrated position for interaction and the perspective the construction is seen from. This is the position the designer needs to keep in mind when developing for experiences will take place in the course of the game. The choice to take a user's perspective is derived from the mimetic theory of narration, a concept introduced by the Greek theatre which concerns how an object is perceived and presented to the beholder [7]. This perspective is also essential in both user-based design as well as in game design [14, 30]. Within digital games Fullerton also uses the term character as the player's dramatized representation but makes a distinction between the character and the avatar, stating that, the avatar is created and formed by the player [14]. Here the "character" does not assert how it will be represented and is only a position for the designer to take the perspective of the user when outsourcing the information (plot).

*World*
The world element represents the space, which the character experiences while moving within. The elements of the world are the inhabitants, contents, etc. From the character's position (the user's) the world interacts with the character and vice versa. The world is the space where choices, explorations and actions occur.

*Action*

The action element is where the action takes place. It represents those inputs and outputs generated by the user's meeting with the world, its habitants and objects that create an action. By knowing the character, taking its position as a user, encounter the world and create a balance of meaning between these enable the receiver to cognitively create patterns and strategies for choices and action.

The three story components, or elements for information, lay a base to mange the interplay between *narrative*, *interactivity* and *spatiality*. The elements or information also represent and invite the creation of the diegetic world system. And through causality, the *logic*, *laws*, *rules* and *relations* can be formed.

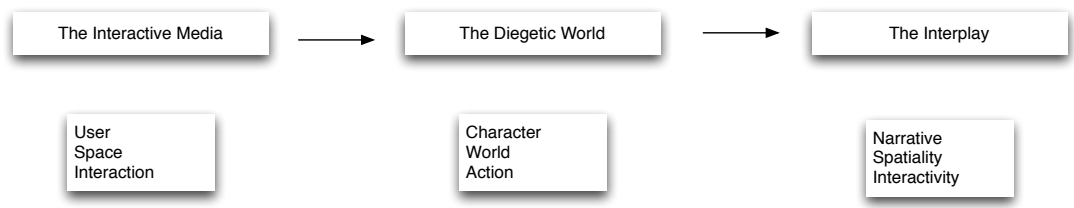

**Figure 1:** How the systems and elements of the interactive media are transferred and represented as a diegetic world to create interplay between narration, spatiality and interactivity.

## 2.3.2 The syuzhet and style as two systems when constructing the fabula

In chapter one it was briefly mentioned why some technique-driven game design did not work. Within narrative theory there is an explanation to why it does not work having a superior technique driven media and ends up as "eye candy" [12]. The explanation lays in the two narrative systems called *syuzhet* and *style* [7]. Syuzhet is the construction of the information (the plot construction) and the style is the means by which e.g. sound, audio, text, graphic and technique support the syuzhet.

When looking at a practise to see how the narrative systems work and further explain the phenomenon there is the example of the agent that was made for a theatre performance nine years ago (see section 1.1) [8]. The task was to make an agent represented as an interactive confessional box "fun". The team had already created an agent they were not happy with and as the agent was already designed, the plot (the syuzhet) had to adapt to the style that was already extant. This meant that the narrative syuzhet system, the plot, had to adjust to the technique, AI, props, colours, audio and mechanics that belonged to that particular system. The scriptwriting became a retrospective work covering design decisions that could not be changed. For example the AI was already set by the construction of the conversation program with a confession in three acts (introduction, confession and absolution). The old agent, a priest, had been designed with three moods: a neutral, sad and angry. The new agent, a rascal and a naughty character, had to fit these moods commiting to receive the confessions and offer absolution. If the syuzhet had been given its proper role and constructed for the new agent, the character might not have conducted the confession. The agent also had to be adapted to have a sorrowful mood, as the sadness was constructed to suit a priest that showed compassion. Even if the work turned out well, it is most likely that the experience of the new character would have been stronger if the syuzhet had been set from the beginning. Instead, the plot had to focus on adapting the new character to fit the template of a solemn priest.

The advantage of setting the syuzhet first suggests the creation of the style system. A game that turned out very well and where a team worked intuitively with the syuzhet and style was "Silent Hill" [25]. Keiichiro Toyama [31], the creator of the game, and Akira Yamaoka [32], the sound creator, were interviewed for a magazine about the noted horror game. Yamaoka, the sound creator, tells how he took a part in the design process from the very beginning in the project – from story to the control of the game – which made him understand how the music could contribute to the whole. Toyama, the creator, said he wanted to make something different, a suggestive game. He had been inspired by the sense of loneliness in David Lynch's "Twin Peaks" and wanted to create another kind of music than the traditional jarring which Yamaoka by his way of setting the music and sound to the story prevented.

Syuzhet and style are simply two systems that cannot exist without each other when constructing the fabula. However, syuzhet is superior to style as they treat different processes and setting of information where the style supports the syuzhet to emphasize the experience [7].

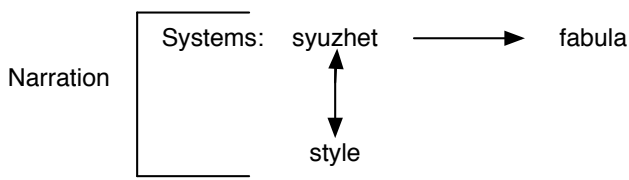

**Figure 2:** Bordwell's illustration of the syuzhet as a system for arranging components and the style as a means of mobilizing techniques. The two-way arrow shows how syuzhet and style interact whilst treating two different processes, the fabula represent the set of inferences [7], p. 50.

Making "Silent Hill" [25] was an intuitive work, which applied the syuzhet as superior to the style system (visual, audio, music, text, special effects, etc.). The diegetic world was designed by setting the objects, their attributes and relations and afterwards the sound was added as a style. (In diegetic theory the style is also called *extra diegetic,* being additional components to the diegetic world but nonetheless valuable for the world's enhancement).

So what happens if only music and sound are used in a system? Then I would stress that these systems presented are based on arranging objects and their attributes to motivate a player to interact. If one were to design an interactive sound game, the sound would have to be arranged in such a way that the receiver might recognize that causality and make conclusions to make choices with or without a narrative.

### 2.3.3 Three syuzhet principles

The diegetic world does not generate itself, so to construct it one needs to set out the information that creates the logic, relation, laws and rules. The system of the syuzhet has three principles for this process: *narrative logic*, *time* and *space*. These principles are used in the plotting of causes-and-effects towards the premise. By taking the information from a premise by asking where, when and how things happen, a relational vehicle between the elements in the systems is created. The syuzhet principles simply tell how to cue, channel and manipulate the information through out the diegetic world and can be described as follows [7]:

*Narrative logic*
This is where the phenomenon of events is arranged and the relations between events are established. Events can be constructed as linear or non-linear by blocking or complicating the construction of relations.

For example, in the film "The Matrix" [33], we meet a normal guy called Neo and become familiar with the world he lives in. The presentation of this world is arranged in such way that the world and its conflict is shown before Neo is introduced to the viewer. The information is arranged so it blocks a rapid construction and through this delay a construction about who is who and who can be trusted. When the main character, Neo, is presented in the film the receiver is as puzzled as Neo when the information about his relation to the world is revealed. In making games one has to think the same way about how the player shall learn about the world.

*Time*
The syuzhet can cue events in any sequence and these can occur in any time span and frequency through repetition. The syuzhet can block the receiver's construction of the fabula – e.g. genres like crime stories in which the fabula is known to the receiver, or in those genres where one has to stay in a predefined cuing of the syuzhet as this is what the receiver expects.

For example, Neo meets his antagonists on several occasions while he is growing and training towards the final victory. Using the principal of time, the constructor decides how many times these meetings will occur. As the film "The Matrix" is already structured according to the canonical story format the frequency is already decided, but in a digital game the space has to be taken into consideration along with how to disseminate and access the information. As if one has an emergent system the frequency might increase to assure the player comes across the information in the "open" space.

*Space*
The syuzhet helps to create a spatial environment by informing about surroundings, positions and paths but it can also hinder comprehension delaying, puzzling, and even "fooling" the receiver's construction of the fabula.

For example, Neo starts off in a normal city that turns out to be chimeric one, which he can leave by picking up a phone, having the person on the other end help him leave the matrix and reach the real world. This is highly relevant for games so the player experiences how to move about in the world, what to avoid, how to access the world, what to control, and not.

## 2.3.4 The premise and "causes-and-effects"

The syuzhet, or the plot creation is often presented as a canonical story format but they are not the same [7]. To separate the two the minimum of data needed to construct a fabula is, according to cognitive research, a *premise* and a *cause-and-effect* according to cognitive research. The *premise* can be of any scale, from a piece of an event, a beat or sequence, to a whole concept. It is a process by which causes-and-effects are plotted towards the premise [7]. Fullerton describes the premise as a setting of time, place, the main character (-s), and the goal and the activity that forward the story [14]. If one takes the film "The Matrix" [33] a premise could appear as follows: An ordinary guy is contacted by mysterious strangers telling him that the world is a computer-controlled virtual world and he is the Chosen One that has to save humanity from the machines. One can see how the premise, even if only expressed in one line, sets the beginning of a construction of the causes-and-effects. There is an ordinary guy who gets contacted by strangers and their unlikely meeting would not occur if it were not for the fact that the main character happened to be the Chosen One, etc. How this meeting goes and what happens are then plotted by syuzhet principles mentioned in section 2.3.3. Within narration the "cause-and-effect" is often related to the creation of conflicts or obstacles. A conflict does not necessarily have to be good versus evil or life versus death. In "The Matrix" a conflict can even be seen in how the ordinary guy, called Neo, is depicted in a contradiction to the mysterious strangers that are rebels dressed in leather and have supernatural ways of fighting. When developing plot (the syuzhet) for digital games, the cause-and-effect need to be seen from the player's perspective as the decision is within the player and it is the player that should carry out the activity. The impact of the effect depends on how the cause is modelled in the diegetic world depicting e.g. Neo's world and how the premise is constructed, and what experience one would like the receiver to have.

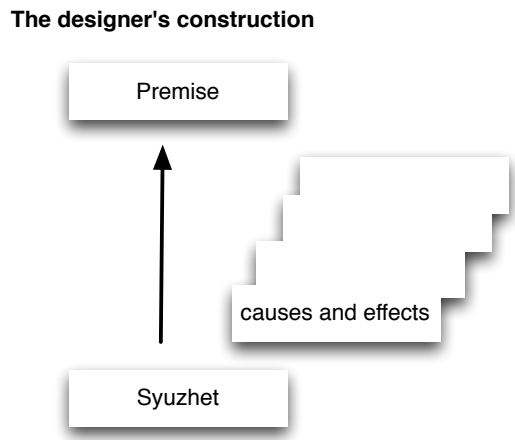

**Figure 3:** The minimum data needed for the designer to create an experience.

## 2.3.5 Goal and "conditions and consequences"

A receiver shall "read" the small bits or sequences of causes-and-effects that have been directed towards a premise. The inference is called the fabula and the minimum needed to motivate the receiver to "read" the information is to provide the receiver with a *goal* [7]. The drive-to-goal in "The Matrix" [33] is that the ordinary guy Neo gets chased by the machines and sees no other way to survive than joining the mysterious people that believe he is the Chosen One. Fullerton [14] refers to the goal as the experience the designer likes the player to have throughout the game and stress on the importance to keep it in mind through every single stage of the development. Seen from the narrative and cognitive perspective the drive-to-goal" is a process that has to be triggered by the designer. Within the digital game it is important to remember to set the goal so the player will be able to carry out the activity, as well as to allow for creating mental patterns and strategies for choices and even empathy if well established. To be able to create the drive-to-goal" and a motive to carry out the activity, it is important to establish the diegetic world and its logic, rules and relations. In the diegetic world of Superman it is possible to wear pajamas and fly. From this, the receiver can draw conclusions and see consequences from their allowing them to learn the conditions of this particular world that would allow them to accomplish goal, such as for instance saving the world. Depending on if the player will be Superman or someone that needs Superman, the conditions and consequences will look different when the player create mental patterns and strategies for choices.

How the premise and goal interact can be seen in "The Matrix" when the causes-and-effects are set so the conditions and consequences leads the ordinary man, Neo, to not only join the supernatural and mysterious strangers but also take the challenge to save the world. If it were a game, one would think in terms of what experiences one would like the player to have. In the same way as the premise, the mental pattern and the experience are set within the diegetic world by the created laws, rules and relations of it in balance with the media- specific attributes of the digital game. If the premise and goal is well-balanced one can see a similarity between them, as e.g. a premise saying that a man is going to save the world and the goal and experience would be a man feeling pressure and stress while trying to save the world.

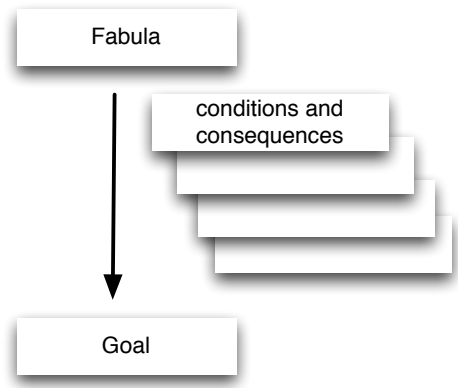

**Figure 4:** The minimum data needed to motivate a receiver to "read" the information and to understand the conditions and consequences in the world in order to reach a goal.

## 2.4 Aligning the interrelating systems and elements

Previously in this chapter the digital game was defined in terms of the media-specific attributes of interactive media as having a user, interaction and space. The main elements creating a digital game were the *narration*, *game play* and *"style and mechanics"*. Using the "massively multiplayer online role playing game" (mmorpg), "World of Warcraft" [5], as an example, the systems and their elements will be put together to show how they provide a base for the method. It will also be explained why the "game mechanics" or "game engines" inspire and form a new term, the *style and mechanics*, influenced by the narrative systems.

### 2.4.1 The Diegetic World

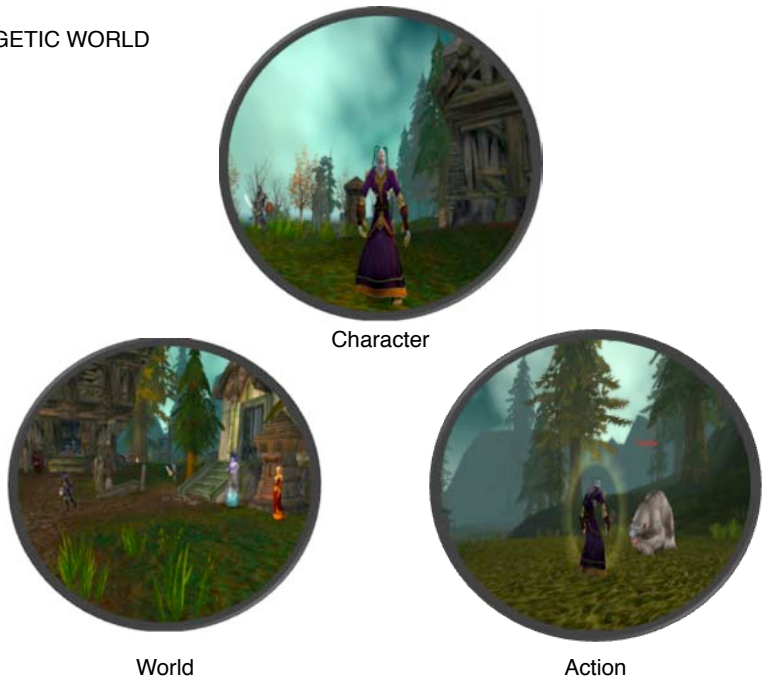

**Figure 5:** The diegetic world showing the character, world, and action.

The *character* is an Undead that is a race one can choose as a player to play in World of Warcraft which has a background story. The Undead are placed in a *world* with inhabitants and a culture and their *action* is to defend the world.

### 2.4.2 The Interplay

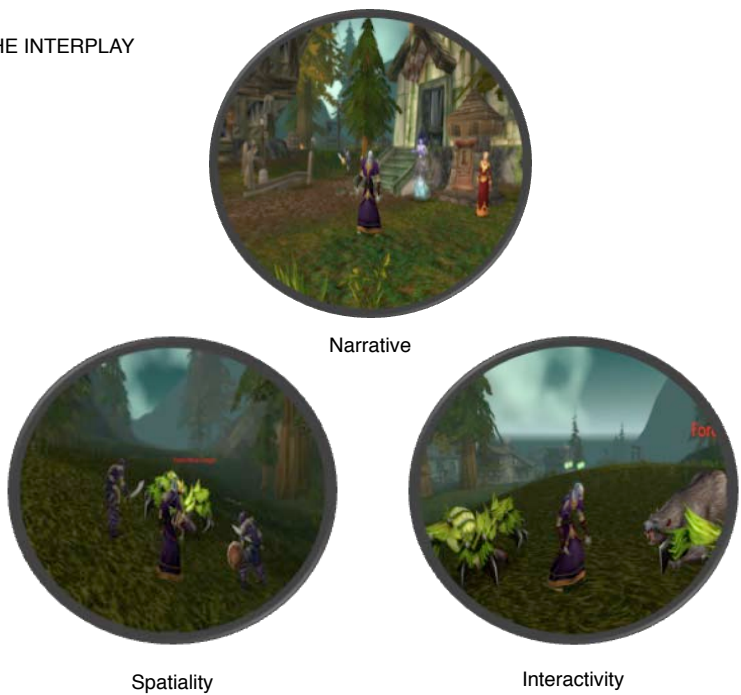

**Figure 6:** The interplay between narration, spatiality, and interactivity.

The Undead needs to defend its world where all elements in the diegetic world are represented in the *narrative*. The *spatiality* tells what and where the threats are in the diegetic world. In Figure 6 the character is shown outside their village, something which is presented in the narrative, while *interactivity* informs how the defence of the village is carried out by fighting, killing, and (hopefully) surviving.

### 2.4.3 The Digital Game

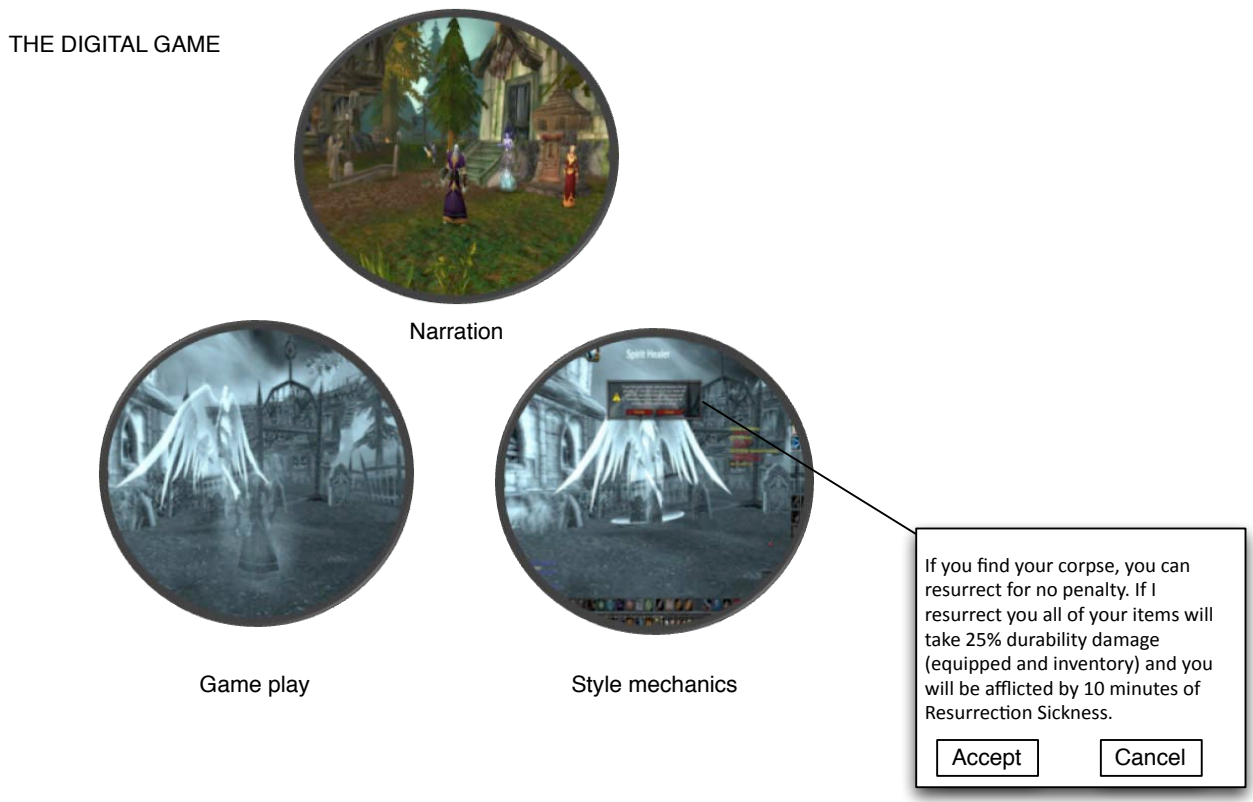

**Figure 7:** The digital game showing the interrelating systems of narration, game play, and style and mechanics that set the game world.

All information retrieved from *the diegetic world* and *the interplay* is here gathered in the element of *narration*. This example shows the systems activated when the character dies defending its world from threats, as well as, how the narrative systems of syuzhet and style are activated (see Figure 2, section 2.3.2).

"Dying" in the World of Warcraft is treated in the system of *game play* and is a part of defending the world thus influencing the player's interaction (see section 2.1 and how game play is defined here as the interactivity formed by what the space offers the player via its narrative). Dying also generates a special mechanism that the designers have chosen to handle the final activities involved in death: the character meets an angel called "Spirit Healer" who, seen from a narrative perspective, offers the character a new chance at life.
In the *style and mechanics* "to die" is visualised by the angel. The death, when related to the narrative system *style*, is described by an image that goes from coloured to greyish, the sound turns into a whispering voice from Spirit Healer and an icy wind can be heard blowing. The *mechanism* of "dying" is an offer given by the angel (see Figure 7) to return back to where one died and recover the body. The player can also give up and get resurrected by the angel. This example also highlights the non-narrative effects, such as taking damage on all gear by 25 % and to become infected by a sickness that reduces the character's ability to fight for ten minutes. Via the GUI (graphical user interface) the *style and mechanics* also allow the player to change their mind using the application's "accept" and "cancel" options. This forms the definition of *style and mechanics*.

## 2.5 Overview of the method and its interrelating systems and elements

Through the previous sections, systems and elements have been presented to form a method, while also discussing its various causalities and interplay. So before describing the working process all the pieces showing the process are here presented (see Figure 8). These are the pieces that enable a narrative bridging towards the interactive media and the digital game. The pointers and the phases represent the iteration that will be explained in the following sections.

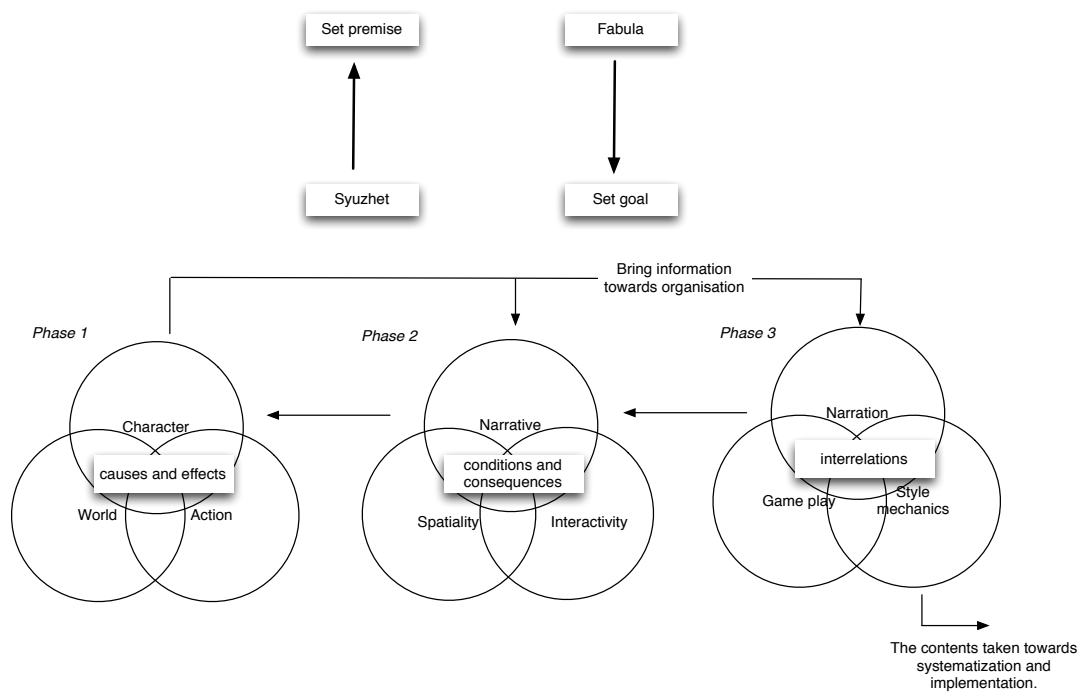

**Figure 8:** An overview of the method and its interrelating systems and elements and the iterative process. The Universe of Discourse for the three Venn diagrams in the figure is not explicitly depicted.

## 2.6 Three working phases

The method is divided in to three phases for the purpose to aid the design process when generating, organising, and controlling the information. This information will grow throughout the iteration. Phases are numbered but this is more for the organisation and does not depict an order for execution. The phases generate new information that can be transferred, back and forth, through the phases and finally be systematized and implemented.

## 2.6.1 The start

First we set the premise and the goal (see section 2.3.4 and 2.3.5). To *set the premise* at this stage is to make a simple one-liner telling what the game is about. This is to inform everyone about what we are trying to establis and construct while establishing the causes-and-effects. To *set the goal* means to explain what the receiver shall experience in only one line. It is important to keep both the *premise* and the *goal* in mind through the whole process (see Figure 9).

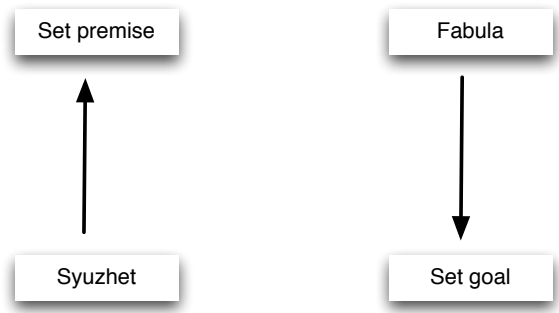

**Figure 9:** Setting the premise and the goal.

## 2.6.2 Phase 1: The diegetic world

The diegetic world shows a storyworld's *logic*, *rules* and *relations*. To plot the logic, laws, rules and relations is accomplished by mapping the elements of information to each other. It is here one also sets the causality, the *cause-and-effect*.

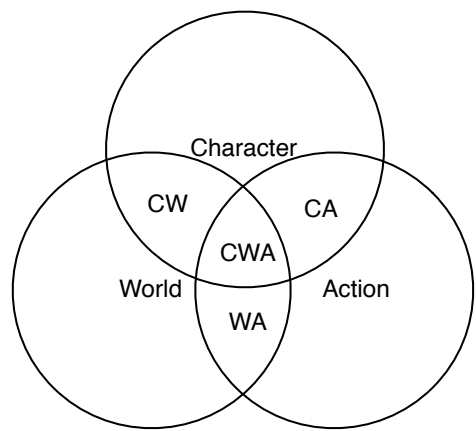

**Figure 10:** The Diegetic World and the interrelating elements of information.

Creating the diegetic world with the aim to create logic, rules, relations, and causality can be described as follows (see Figure 10):

1. Fill in the information you have that fits in to the categories: Character, World and Action.
2. Look at the *cause-and-effect* between the values you have set to the Character and World (CW) and keep premise and goal in mind.
Continue this action and look at the *cause-and-effect* between WA and CA to generate a larger picture of the diegetic world. The diegetic world, and the causes-and-effects, are revealed *by asking who, what, when, where, and* most of all, *why*?
3. Compare how the Character, World and Action relate to each other in CWA to suit premise and goal.
4. Iterate the components and start over again adding the new information generated by the process to the elements of information.
5. Continue iterating until you see the diegetic world complete enough to suit the premise and the goal.
6. Continue iterating until you see that the diegetic world has no loose ends and that logic, rules and relations (meaning, sense) are created.
7. Move the information to Phase 2 to generated, distribute, organise, and control new and further information.

### 2.6.3 Phase 2: The interplay

The interplay between narrative, interactivity and spatiality can be described as the causes-and-effects between the elements of information in the diegetic world. In this phase one finds out where to start, where to go, what to meet, when to meet, how to meet the world and its inhabitants. It is in this phase one can see if the laws, rules and relations, constructed in the diegetic world (Phase 1), are developed in a way that they create clear conditions and consequences for the receiver. If the diegetic world is elaborated it shall be possible to generate further information by working with the suyzhet principles as logic, time and space to see how the causes-and-effects generates conditions and consequences to motivate the user to take action. This phase will generate more detailed information by asking "how", "when" and "where".

THE INTERPLAY

NS = Narrative and Spatiality
NI = Narrative and Interactivity
SI = Spatiality and Interactivity
NSI = Narrative, Spatiality and Interactivity.

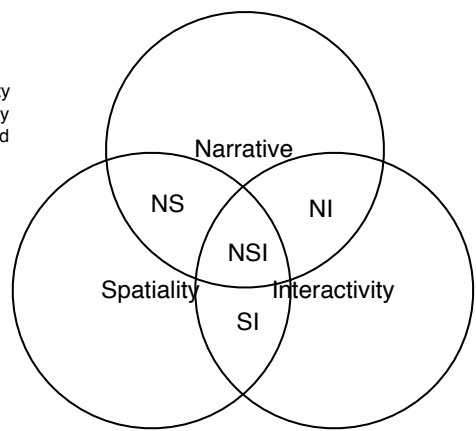

**Figure 11:** The interplay between the elements Narrative, Interactivity, and Spatiality.

The working process:

1. Keep the premise and goal in mind throughout the entire process (see Figure 9). Bring the information from Phase 1 to this phase and focus how the narrative (the construction of the diegetic world), is experienced from a users perspective.
2. Move through the elements Narration, Spatiality, and Interactivity (NS, SI, NI and NSI) and see what is generated when asking:

   Where?
   Think about the spatiality in the environment and where the information about surroundings, positions and paths are. What information would you like to delay, hinder or puzzle for the user and what shall be pushed forward as the first information to encounter?

   When?
   Think about the time and cuing of the information between the Character, World and Action. When shall events occur, in what frequency and repetition?

   How?
   Think about how the information shall be forward and established between the narration, spatiality and interactivity to the user. How shall the receiver get to know and learn about the world and its habitants? Events can be constructed as linear or non-linear by blocking or complicate the construction of relations. How do the blockings and the complications appear?

3. Through this process details will be generated that one might not have detected in Phase 1. Bring these back into Phase 1 and add this new information to the elements of the diegetic world. This process will generate more information that will then be move back to Phase 2. Circulate the information until a result suits the desired premise and goal. If the simulation turns out to not work, see if the premise and goal has to be changed, when restarting the iteration from Phase 1.

4. Find out what kind of causes-and-effects that create conditions and consequences that motivate the user to create patterns and strategy for choices and take action upon. This is made in the interplay between Narrative, Spatiality and Interactivity. Move the new values to Phase 3 to set up a system for game play.

### 2.6.4 Phase 3: The digital game

In this phase, the three systems of *narration*, *game play*, and *style and mechanics* meet and the work is to organise, distribute and balance the information retrieved from the first and second phase to form the digital game. This is to prepare the material that is ready for systemization, scripting and implementation. As said before, the game play is here the interactivity that is formed by what the space offers the player via its narrative (see section 2.1).

The information from Phase 1 and 2 is gathered in *narration*. The *action* and *interactivity*, retrieved from Phase 1 and 2 have been differentiated when modelling the diegetic world and the interplay. In Phase 3 they will here present a pattern of activities in the world for a game play. In other words, the conditions and consequences can be seen here, and if one cannot see and set the game play in Phase 3, one needs to move back to Phase 1 and 2, to see what has been missed in the diegetic world. If Phase 1 and 2 is thoroughly balanced by establishing a logic between its objects and attributes, the causes and effects will be so clear in Phase 3 that a pattern will be seen to set a game play. If seeing a clear pattern, this also means that the player can create strategies and make choices from the developing of the information in the diegetic world based on the premise and goal.

In Phase 3, the organisation of the information in between narration and game play, will generate the possibility to enhance the narrative, game play and desired experience for the user. This is made by setting the *style and mechanics* because one now has detailed information that enables the setting of sound, music, graphics, to develop the user interface, etc, (see Figure 7, and section 2.4.3 about Style and mechanics).

THE DIGITAL GAME

NG = Narration and Game play
NS = Narration and Style mecanics
GS = Game play and Style mechanics
NSI = Narration, Game play and Style mechanics

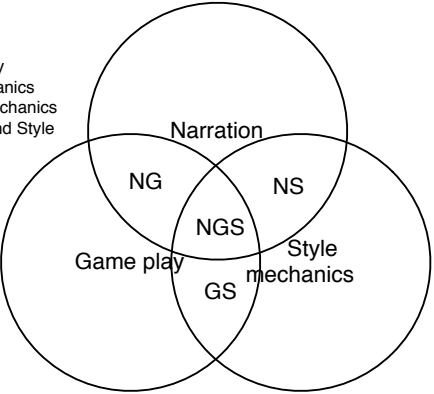

**Figure 12:** The digital game and its interrelating elements Narration, Game play, and Style and mechanics.

## 2.7 Quick reference guide for the method

Here is a quick reference guide for the method.

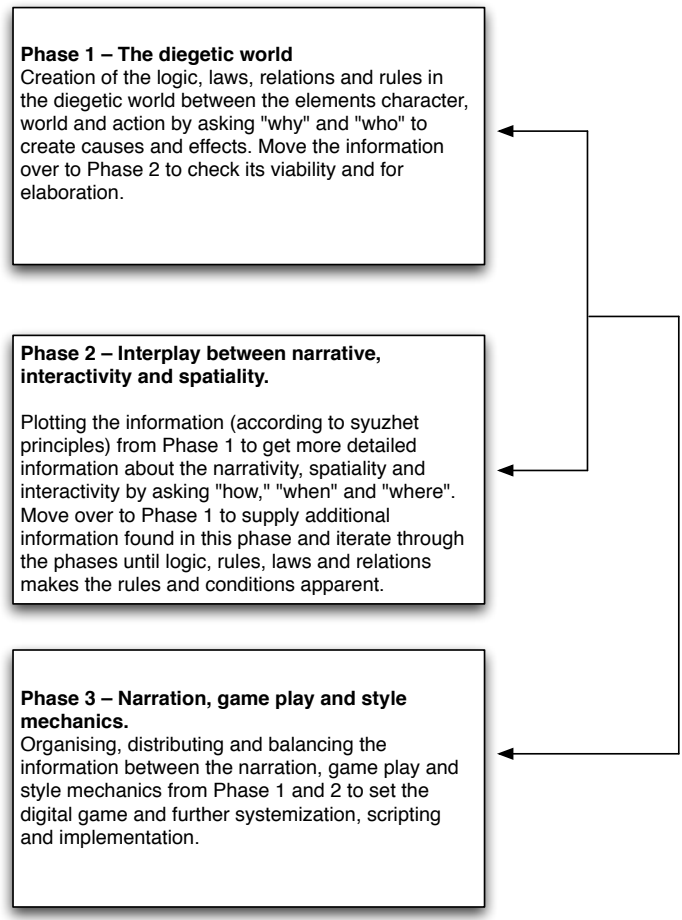

**Figure 13:** Quick reference guide for the method.

# 3 A qualitative study of six game ideas

## 3.1 The preparation, advices and challenges

Methods often arise from the lack of a tool and then end up becoming a tool just for the inventor. It is true this method came from being short of helpful tools when analysing and trying to explain narrative mechanism. When seeing that the tool could be useful for others, helping communicate about narrative in design and provide designers a possibility to consciously work with narrative process, a work began to offer the method to others as it helped me. The choice to try out the method on digital games was the usage of narrative, and the entertaining capacity within the games. It was also more likely to find people with experience in narrative that could try the method.

To find subjects for the study required a bit of thinking and advices. The task should be to present an idea of a game, iterate the idea through the method and then compare, and discuss the result. A seminar and instructions were also needed. The problem was finding someone that had the time within the game design business, someone very unlikely to find. Even students with a tight schedule would be difficult to attract. The advice I got was to find a course at one of the universities teaching game design and within that course give a lecture and do a workshop. At that time a university was about to start a course in advanced design for their game design students and was going to try out different methods through rapid prototyping (e.g. MDA which stands for Mechanics, Dynamics, and Aesthetics [18]). The preparations to do the study were hurried. As there was no time trying the method on a pilot group, the method was tried out with the help of Emanueledomenico Sangregorio, graduate in Computer Science, at Catania University. He had a similar background as the subjects, and through modelling a game idea and applying the method to this work his questions helped me calibrating the method and how to explain it. The questions he asked were about the phases and how to move the information around and what the different outcomes should be in the phases. Finally the biggest challenge was how to teach my over ten years of experience in narrative practise and present a method during such a short time.

### 3.1.1 Set up, task and schedule

Two weeks were devoted to the study to try out the method at a university course in "Advanced design" and "Rapid prototyping", a requirement for the game development program. The idea behind the planning was to give the students a chance to reflect upon the narrative before I presented the method. The wish was to develop interplay between narration and game play and how the systems interrelated. The students were asked to hand-in a written game idea that should be elaborated. They had one week to complete the task and were given the theme "narrative". They were asked to give a detailed description of a certain part of the game showing how narrative and game play interacted. For the oral presentation they could bring pictures that could give an idea about the world, the avatar/character, and the interaction. The students were also asked to work in groups of three. The students got the task from their main teacher who also added that they should work with the interplay between narrative and the game play. The subjects recieved my e-mail address so they could contact me if they had any questions. During the first week no one contacted either the main teacher. The students were aware that the method was going to be presented and had not been tried out and offered to a research project as an instructional aid.

After one week I met the students for the first time. Six groups were formed and six ideas were handed in. The students got further information about the method and were asked whether they would like to participate with their work and opinions. Five groups volunteered. The sixth group considered they had too much work but joined in the end of the study. I presented myself, told them briefly about my background and afterwards the students presented their ideas orally. At the end of the seminar I gathered all ideas and prepared for the next day when I should give a lecture about narrative construction, the ideas behind the method, and afterwards simulating, with the students, their game ideas with the help of the method.

The second day I started with handing out a questionnaire to find out what the subjects' knowledge of narrative along with their thoughts about the narrative. The reason was to know how it had influenced the game idea and how to develop the topic (see section 3.3) before I started the lecture. The questionnaire was also to see how to teach narrative within game design. An intense lecture was held and the designed ideas were simulated through the method. It should be added that the last question in the questionnaire was related to their game idea and how they came to think about the idea to be able to evaluate the whole process (the setting of the premise, or goal or both).

The following sections in this chapter will present the subjects responses to the questionnaire, their game ideas, the simulations and the iterations together with the responses to the outcome of the iteration and use of the method. This would show the changes they made to their ideas and tell about the process. The last day I met the groups, I decided to see the groups, one on one, as I understood that they had faced some problems with the limited time. The more private discussion with the groups provided a depth to the design situation and the decisions they had made and how the method had worked for them.

*The schedule in short:*

**SCHEDULE**

Day 1  Students get the task to elaborate a game idea.
Day 2
Day 3
Day 4
Day 5
Day 6
Day 7
Day 8  Oral and written presentation of game idea.
Day 9  Lecture, workshop and presentation of the method.
Day 10 Students get the task to iterate the game idea with help of the method.
Day 11
Day 12
Day 13
Day 14
Day 15
Day 16 Oral and written presentation of the iterated game idea.

**Figure 14:** The course schedule.

## 3.2 Views, opinions and ideas about narration

Twenty students attended the course and study. Nineteen males and one female participated and they were between 22 to 29 years old. Fourteen questionnaires were handed in and some chosen to answer together about how they viewed narration. The questionnaire was handed out before the lecture to get an idea about the students' opinion and knowledge about narration.

The *first question* was about what function they thought narration had. All of them replied that narration was to tell a story, two said to deliver a message or an experience. Three of the replies referred to narration as a construction, plotting and technique. Three of them related narration to create emotions and one said narration helped the player to understand the background of a character and the motivation for what they are doing and why.

The *second question* was if they saw any *disadvantage* with narration and the answers were:
"Conflicts between narration and game play".
"Bad when you have to listen to the story and not play".
"It takes time from the developing".
"If the story is badly written the player gets uninterested. If the game does not need a narrative a narrative can feel forced upon the player".
"It risks the player being forced to have a feeling".
"If the narrative is badly written it can stand in the way of the game play".
"Cannot see any disadvantages with the narration unless it is badly used and too abstract that can confuse the player".
"A vague narrative can make the player uninterested".
"The narrative can remove the focus".
"No disadvantages, unless it steals time from the developing of the game mechanics and the game play".
"If it consumes too much time and dialogues are badly written".
"Can create a static end and not let the player creates their own story".

The *third question* was about the *advantages* with narration:
"Narration can enhance the game play and motivate the mechanics depending on how the game play is defined".
"It helps to actively make choices and control the story to be shown and not shown".
"Can strengthen game play".
"Emotions. Emotions drive the human and are our motivation to many things. Narration can make us emotional. It can tell things so we understand new perspectives".
"It enhances empathy".
"Creates an interest and increase the immersion".
"In a war scene narration can make it more believable"
"Narration can give the player a purpose and goal to what he does. It also helps the player to feel involved in the game world".
"Yes, if well-made the story will be the reason for playing".
"It works like a carrot that makes the player move onward".
"A good narrative nails a player to relate to the character and understand why they act as they do and feel".
"Enhance understanding and empathy".
"Creates a depth and emotional connection".
"Create depth, empathy for the characters and develop a message".
"A ramification for the creator to understand the working space".

## 3.3 Presentation of six game ideas

The different ideas will be presented as follows:

1. *The game idea in brief*
   The group's idea will be presented in brief based upon the documents handed in and upon the oral presentation.
2. *Simulating the idea with the method*
   The simulation of the idea is shown as graphically to visualise how the elements in the system is organised. The organisation of the elements and the three phases will be explained in text as well. When doing the simulation during the seminar I thought it was important not to interfere with the group's intentional ideas. When the method detected missing or unclear informations the group was to given free to decide what to do with the result of the simulation and decide wherein to make changes or not. My role was to be a guide for the method.
3. *Results of simulation and iteration*
   The iteration of the idea will be shown with a graphic to enable a comparison of the original material and the iterated. The changes are shown with bold fonts to enable a comparison between the two iterations.

It should be added that I considered making an appendix of the ideas, as it might be hard to read through the six game ideas. This was not possible as all ideas had something of importance for the result. They also offer an insight about a design process and how to model an idea and its contents that I believe any reader, working with design and creation, can recognize and learn from.

## 3.4 Group 1

### 3.4.1 The idea in brief

The premise was to make a historical game about the assassination of Gustav III at the 1792 masquerade in Stockholm. The goal for the user was to socially interact (bow, salute, flirt, etc.) with the environment, to find out who was guilty of murder. In the real historical drama about the assassination of Gustav III, Ancarström was arrested and executed for the murder. It was not clear how many were involved or who the leader behind the crime was. Inspired by this unsolved drama it was up to the player to find out through socializing and interacting with the royal court what happened before the murder at the masquerade and find the assassin before it was too late.

The pitch was well formulated and the descriptive pictures gave a good idea about the world. The group used keywords to express the world: "fox and geese", "Russian war", "Gustav III", "character development", "party", "gossip", "costumes", "coquetry", etc. *The platform* for the game was a consol with accelerometer for Playstation 3 or Wii to support the game play of social interactions using different kinds of greetings. *The environmental* surroundings were the castle filled with known persons from the political scene, artists, royal court people, etc. The drama took place in a traditional environment such as a ballroom, dinner room, parks, theatre, etc.

The group set up a *narrative structure* that the game should take place in a time span of 24 hours. The player should wake up in the morning (or night), told by a butler that the player was expected to meet X. They presented the player's position dramatically by addressing the reader of the pitch with "you": "Dazed, you are lead through corridors to finally meet X who tells you about a conspiracy against the king and they expect you to find out who is behind it". The player should gradually get a picture of the conspiracy through the game and be given a chance to change the outcome if the player finds out who was guilty. Through the setting, the player should slowly understand what social class they belong to. The environment and people the player meets should tell something about the player, to the player. The playing should mainly focus on socializing and developing an idea of all possible motives to finally make a conclusion and hopefully find out before the assassination happens.

The group explained how the avatar's perspective would be changed through three acts. The player started with a first person perspective to become a third person perspective – a way for the user to grow with the character. The player's main interaction was to socialize and solve the crime by investigating, exploring, talking to people, exchange information for the player's favour and manipulate people by facial and body expressions.

### 3.4.2 Simulating the idea with the method

→ Premise/Syuzhet – The murder of Gustav III
← Goal/Fabula – Socializing to find out who murdered Gustav III

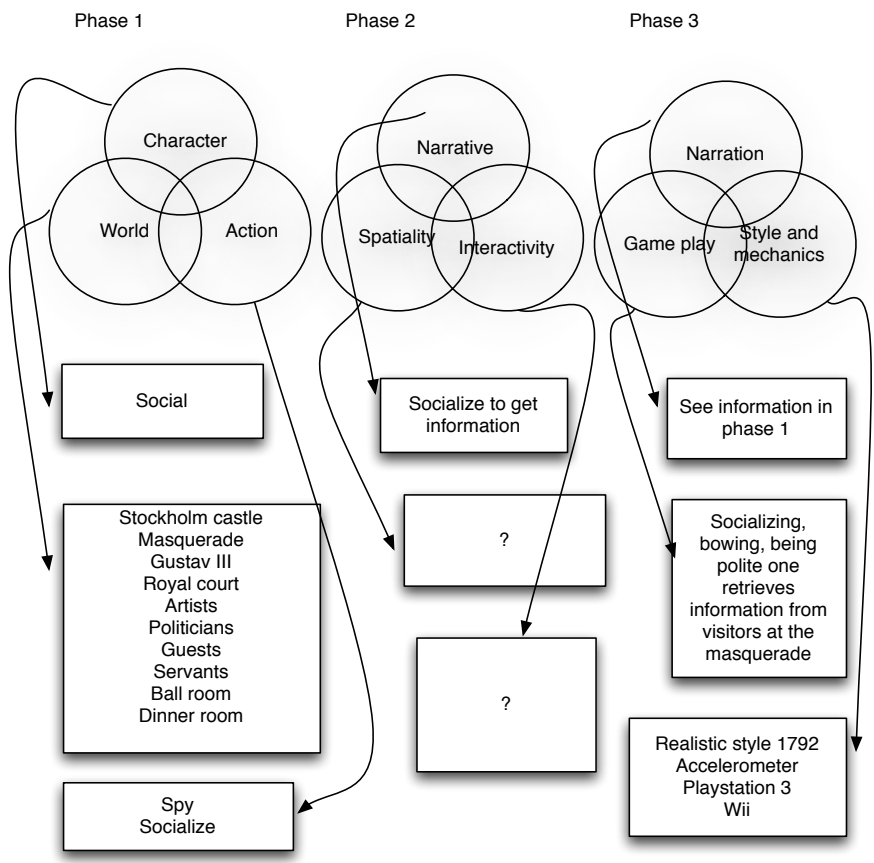

**Figure 15:** Group 1 and the first iteration of their idea.

*Phase 1 The diegetic world*

The group developed the rules, relations and logic to the diegetic world by choosing that action take place in the castle and at the masquerade. They thought about all participants that could be present in the world and how to socialize. They used socializing gestures for the user's action to find out about who lay behind the murder of the king. The character was a person that should socialize.

*Phase 2 Interplay between narrative, interactivity and spatiality*

When transferring the information developed in Phase 1 to Phase 2 to find out where, how and why the action took place the idea faced problems. When looking at the causes-and-effects for the user's spatiality, and interactivity a question arouse – who was the user? Uncertainties arouse when asking how the user moved and where and when to find the information about the imminent assassination. Due to the player's goal (fabula), to socialize through the world to find out about a crime, the question was how they wanted to set this interactivity? Where could the player walk? Could the player be near the king? How did people socialize and pay attention to the user? Did the player's character have a strong position or should it be set as neutral (a visitor from the future just peeping into a historical documentary)? Simply expressed, the user's representation in the drama had to be set to know how to work with the interactivity and the spatiality in the world to know how to construct the access to the information.

*Phase 3 Narration, game play and style and, style and mechanics*

The group had thought through a game play (the socializing) and how to support it with different technical devices as accelerometer, Playstation 3 or Wii. The group also presented a scripted structure for where the event should start, the duration of the drama, and acts as well as actions.

### 3.4.3 Results of simulation and iteration

After the first simulation an irregularity was found and the definition of the user's position. It was for the group to decide what kind of game, premise, they wanted and what goal they wanted the user to have. The group produced a strong structure already with three acts, and an idea about how to begin the game, but needed to elaborate the idea to detect what the spatiality and interactivity looked like. This was to enable game play that motivated the player to take part, know how to take part, and to be successful. The group iterated the idea and presented a solution to the first problem that looked as follows:

→ Premise/Syuzhet – The murder of Gustav III
← Goal/Fabula – Socializing to find out who murdered Gustav III

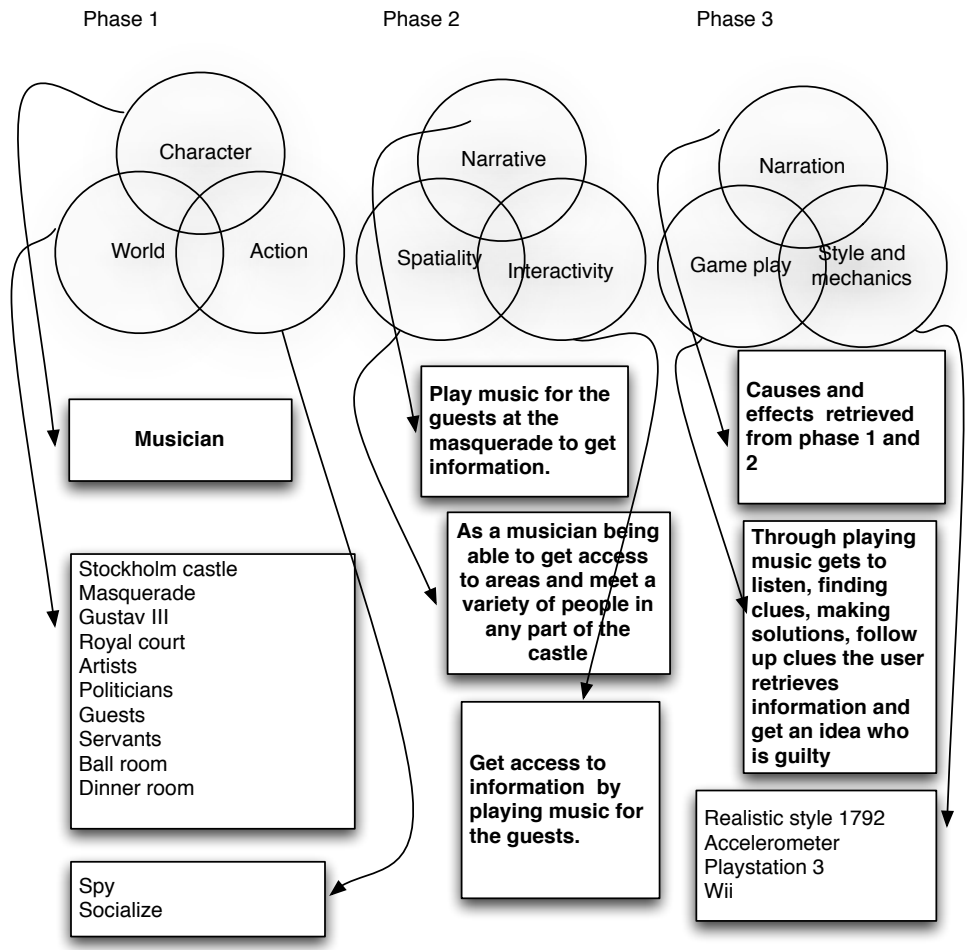

**Figure 16:** Group 1 and the second iteration of their idea using the method.

The problem had been to decide what kind of game they wanted to make to be able to select what the user's position should be represented by. Initially the group wanted the user to slowly grow with the character but this caused problems. The group's choice to use a musician (see Figure 16) allowed them to keep the original premise and goal. Using a musician enabled the group to go deeper into the construction of the spatiality and retrieving interactivity in Phase 2 and try out to see how the socializing and retrieval of information (the spying) for game play could look like.

They had chosen to narrow the spatiality to a dinner room to find out how the interactivity would look like to explore the system for a game play. They gathered eleven people in the dinner room to see who would attend the dinner to find out how and who the user, the musician, could manipulate to retrieve information. This revealed more details to a possible game play by having one missing guest at the table, letting the user hear about the missing guest and how the user could choose to follow up this way, they could find new traces about who the murderer(s) could be.

The subjects thought the method pointed out where the idea lacked flow and helped them to find the problems so they could fix it. The group regretted that the time was too short and they would have needed more time for their work, while also learning more about narrative and using the method.

## 3.5 Group 2

### 3.5.1 The idea in brief

This idea was called "My Poetic Collection" and the premise was about a seven years old girl with a personality disorder, having dark dreams and expressing her world through writing poetry. The goal was for the user to experience the girl's perspective. The group wanted to talk about a girl that tried to save her world, but through her dreams she was pulled towards her dark inner place that she had tried to avoid. The world she met was full of threats and people she met, she killed, and afterwards she wrote a poem about it.

The idea was described as a literary text:

"While the boys stood idly by, their eyes filled with disbelief at what had happened to their friends Angelica swung the knife at the fat boy cutting through the thin cloth of his coke stained T-Shirt effectively gutting the boy. She turned to the skinny boy with a calm almost sweet expression on her lips, his knees were shaking heavily and on the ground beneath him there was a puddle, a big wet spot could be seen on the front of his gray slacks. She took a step towards him and his knees gave away as he started to sob silently.

*Little Boy by Angelica Fenland*
*Little boy,*
*Born alone.*
*This is not your story,*
*Nor is it your song.*
*A world without wrong,*
*Is a world without right.*
*It is an ode to what should have been done,*
*In a night without hope and light.*

*So little boy,*
*Stand and fight!*
*Little boy…*
*This is good night.*

Her hands were bloody as she stood over the corpse of the skinny little boy; it had taken almost half an hour for the screams to go silent, she had hit him so many times yet he wouldn't go silent until she the tire iron laying on the street next to a wheel-less car had crushed the side of his skull."

The group saw the idea as a linear storytelling showing two worlds with innocent victims, the girl and the boys, where the girl saw everyone as hostile and dangerous. The girl tried to get revenge for the imaginative violence towards her, in contradiction to the "real" world, were people just acted normal. The group showed animated pictures of cartoons in a realistic and modern age. The game was seen as a 2-D platform single player game where one runs and kills people.

### 3.5.2 Simulating the idea with the method

→ Premise/Syuzhet – A girl with personality disorder dreams and writes poems about her world.
← Goal/Fabula – Exposed and threatened.

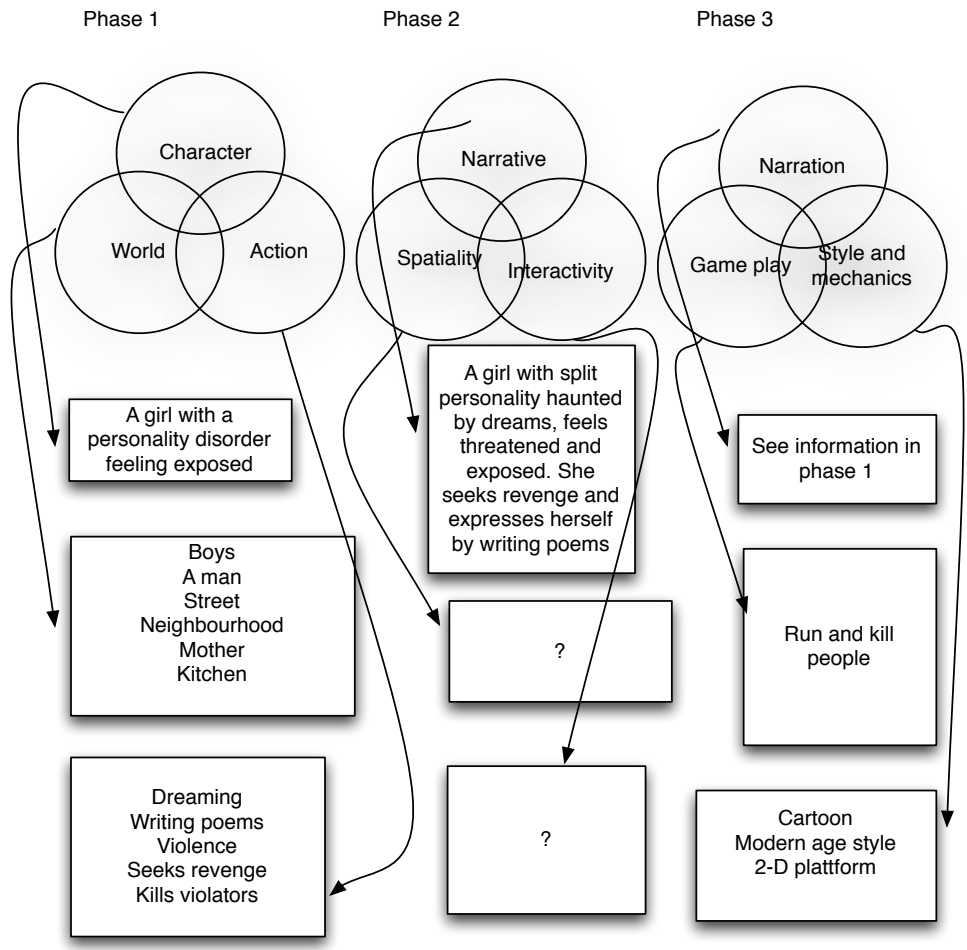

**Figure 17:** Group 2 and the first iteration of their idea.

*Phase 1 The diegetic world*

The group had chosen to describe the game in a literate form and the work was to see how the information could be transferred to suit the interactive media i.e. find a form that allowed for interaction. The relations, rules, and logic were set to a girl in the user's position. A picture was brought by the group to give a visual idea of the world. By reading through the text one could detect a world where the inhabitants were boys saying something about a neighbourhood, a man, and there was a mother that she talked to after killing a boy on the street.

*Phase 2 Interplay between narrative, interactivity and spatiality*

When looking at the causes-and-effects from the information created in Phase 1 and by moving it to Phase 2 (to the spatiality and interactivity) it felt unclear who, where, why, and from whose perspective the action took place. The reason for this was a user's position and a split personality. The group had to decide whether they wanted to change the goal for the user (feeling exposed) if they wanted the user to be motivated to take part in the actions or if they just wanted to stay with an idea with a girl that killed anyone that came in her way. If the group wanted to stay with the existing goal for the user to feel exposed and threatened, the group would have to explain the two worlds that were represented as the girl's world and the real world and the killing of innocent people. If the group would develop the worlds they would get more information about how to discuss the girl's action and to manipulate and queue the information in a way so the user would get a stronger identification with the girl. This would also generate a more clear cause-and-effect. Another important part of the idea was the dream and the poem. If the group develop the world/-s they would be able to find patterns for conditions and consequences to take form. Did she dream first or write the poems, or vice versa, before killing and how did these relate between spatiality and interactivity?

*Phase 3 Narration, game play and, style and mechanics*

As Phase 1 and 2 were not clear one could not detect conditions and consequences between a narration and the game play. The style was clear through the images brought by the group, a cartoonish style in a modern age that showed the "real world" but not the girl's imaginary world. The group expressed the idea as a 2-D graphic.

### 3.5.3 Results of simulation and iteration

The group had chosen to make a game from a literary text and had a big challenge to motivate the user to take action (if they wanted to stay with their original). To do this the group had to define the different worlds they wanted the user to experience. As the wish was to express the girl's action, it was up to them to see how to organise the information. When simulating the idea through Phase 2, following questions aroused: where did she live, how did she move in the world, who did she meet, what triggered her action? The group simply had to define the girl's imaginative world and how it corresponded to the real world to show the logic, and causes-and-effects for the action to create motives. By defining this it would be easier see conditions and consequences for the user to act upon.

→ Premise/Syuzhet – A girl with a personality disorder dreams and writes poems about her world.
← Goal/Fabula – Exposed and threatened.

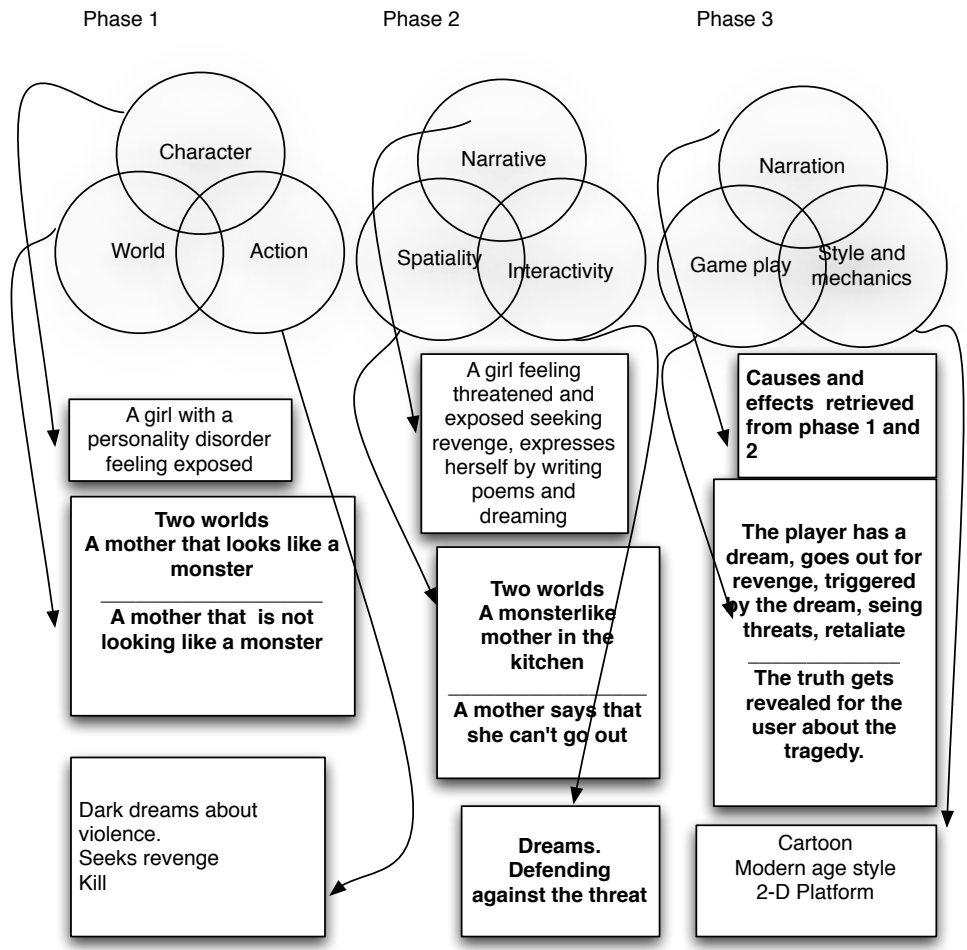

**Figure 18:** Group 2 and the second iteration of their idea using the method.

The group came back after a week of iteration and had kept the premise and goal – the user should experience the girl's exposure and personality disorder. They had challenged a hard task to transfer a literary text to another media. They talked about how they had structured a system for how the dream worked as well as the world seen from the girl's perspective. It should be revealed for the user what the girl (the user) actually had done. The group had chosen to narrow the world and looked at interaction between the mother and the girl to detect a structure. They developed a structure for game play: the girl had a dream, affected by the dream she acted out the message from the dream and afterwards she wrote a poem. Who she ran into and corresponded to her dream became the victim and formed the game play.

What the group said when discussing the idea and how they expressed the game in the text document showed ambivalence to the material. In the document they expressed the iteration and improvements saying; "this character is fucking sick in the head and sees things that does not exist", "the first boss is her mother that looks like a monster", "she kills random people she meets", "what you do is to kill everything you see and solve simple puzzles on the way", "when mother says you can't go out the girl sees it as a physical attack and reacts by killing her mother and afterwards she writes a poem" and "the story is linear and you have no choices". The group thought narrative was hard and the method was hard to use. They felt they would have needed more examples to be able to see how narrative worked and they found the time too short for the iteration. This group was also the one when answering the questionnaire about how they looked upon narration that narration was linear and a hindrance when developing games. The theme of the course "Narration" might have influenced them to take this track as well.

## 3.6 Group 3

### 3.6.1 The idea in brief

The title for this game was "Life begins as an egg". It was a single player game, and the premise was about a parasite that takes over a body for its own survival and prosperity. The goal was to infect and effect the body. It all starts with an egg. To grow, the egg needed to infect different parts of the body, which affected the carrier in different ways. The group said they wanted to create two narratives, the life of the parasite and the carrier's life, effected by the parasite.

The *game play* was to take certain parts of the body and create different reactions for the carrier such as stress, love, happiness, etc. The game had *two levels* that were divided into two parts. The first level and first part was for the egg/parasite to grow and take over one part of the body which the parasite made to headquarter. From the headquarters, the second part, the parasite sent out "commando parasites" that were equipped with different skills and amor. The equipment was bought with money earned via "parasite points" that one received when taking over certain parts. To reach the second level one had to gain control over the body parts and from there it was time to control the carrier's action and reaction.

The *first level* had a first- and a third person's *perspective*. The player should be able to move in a three dimensional way such as up and down, sideways, etc. One moved around in an open space, while some areas were closed by enemies, while others were the immune defences represented by white corpuscles. All levels should have control points that should be difficult to find. This offered exploration, and of course danger, and some points should have stronger defences than others. The second part was seen from a third person perspective controlling the carrier and having access to his life and seeking what the carrier was about to do in his life. At the oral presentation the group also talked more about the carrier. It should be a man in the prime of life, having for example a date with a girl that the parasite causes problems with.

The group presented pictures from games like "Spore" that showed a germ's life from being a seed to growing into a full life form. They also showed a picture of a marriage from "Sims" and a picture of person with a cold, and space pictures of spacecrafts shooting at each other.

### 3.6.2 Simulating the idea with the method

→ Premise/Syuzhet – A parasite that tries to take over a body in order to survive.
← Goal/Fabula – Infect, control, and defend to survive.

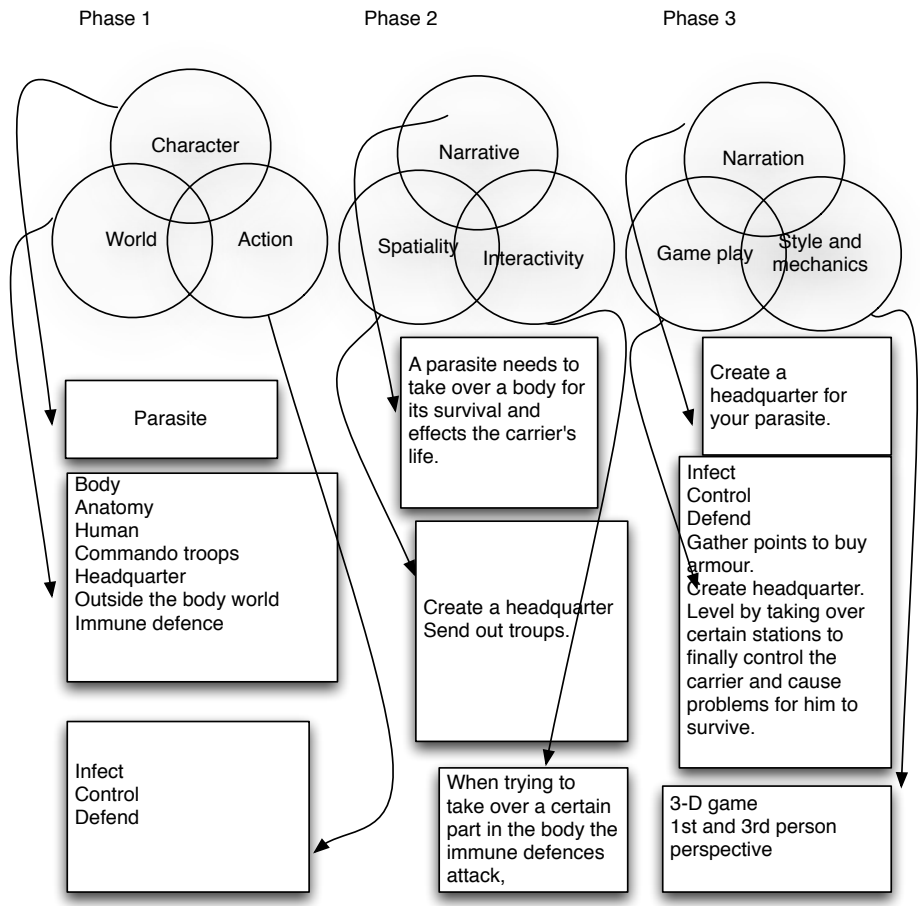

**Figure 19:** Group 3 and the first iteration of their idea.

## *Phase 1 The diegetic world*

The group developed the rules, relations, and logic for the diegetic world by choosing the action to take place from a parasite's point of view and for its survival, and complete different tasks in the game. The premise and goal were clearly defined as a parasite that wanted to take over the world for its survival and for the user to infect, and get in control of a body. The logic, relations, and rules where balanced and created a clear picture of the diegetic world.

## *Phase 2 Interplay between narrative, interactivity and spatiality*

When moving the narrative information from Phase 1 to Phase 2 to find out where, how and why the action took place, the two narratives the group had in mind, were not described. The group referred to a parasite's relation to the carrier and, through visiting certain parts of the body, it should effects the carrier's life. One thing the group had done was to define the setting of the parasite's headquarter, but the directions and the targets and their relations to the carrier had to be elaborated if that was the game they would like to create (the premise). By defining the "two worlds" (the carrier and the parasite) in Phase 1 and seeing how the systems related, the group could see how the causes-and-effects looked like to created a system for a more developed conditions and consequences that formed the game play.

*Phase 3 Narration, game play, and style and mechanics*

The group presented game play showing a parasite and a body. But as the targets, the relations between the parasite and the carrier was not cleared out, the process had to move back to Phase 1 to divide the worlds (the carrier and the parasite). By running the new information from Phase 1 through Phase 2 it would be easier to structure where the parasite was born, what was the first target, what was needed to be defended, what was next target, when did the target need the troops and what were the different effects on the carrier and when or what should the parasite do or use to gain maximum on the carrier, etc? Even if the group would stay with a simple graphic or a simple narrative (not involving a carrier) the group would still need to decide these things to know the targets for the parasite.

### 3.6.3 Results of simulation and iteration

When the group came back they had chosen to stay with the old premise and goal to make a game with a parasite that should infect and take control over a body. What the group had done was to line up the two worlds, the parasite and the carrier.

→ Premise/Syuzhet – A parasite that tries to take over a body and teach a person to become a better man.
← Goal/Fabula – Infect, control and defend to teach a man.

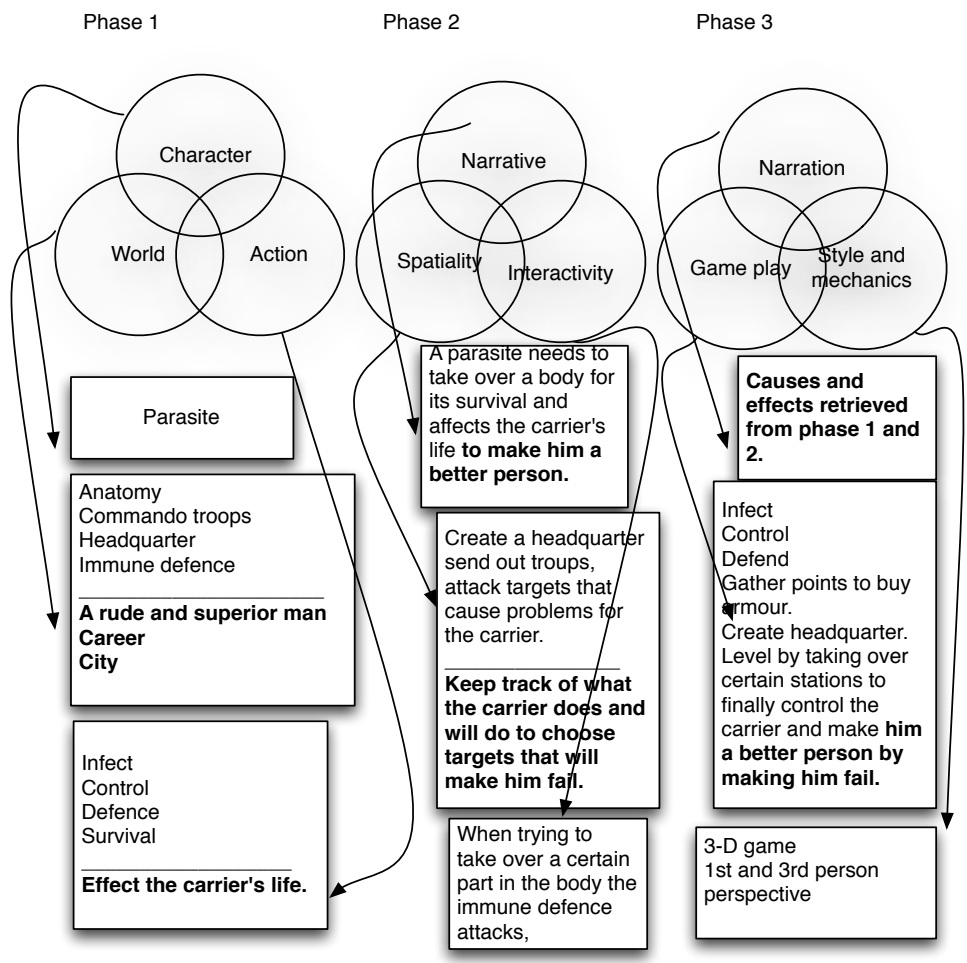

**Figure 20:** Group 3 and the second iteration of their idea using the method.

What happened after the iteration was that the parasite's goal to survive and take over the body got an extra layer and motive for the player. From the beginning it was about a parasite's biological concern to take over a body and to effect a person's life. After the iteration the group had developed the carrier's world and made him a rude person at the top of his career that they wanted to see fail (learn a lesson). What the group also did by elaborate the worlds (the parasite and the carrier) was to give a depth to the game play. The group could have chosen to work with game play based on a system where the heart and brain would have been the hot spots for the attacks. Instead they related the targets with what the man was about to do, in order to let the parasite attack more subtle systems in his personal life or career. The group had to little time to develop a full game play system but they had seen new possibilities too like making the carrier sneeze, fart, have involuntary reflexes, etc, when seeing a girl or attending a board meeting. All information was found in the first idea but appeared better after the iteration.

The group thought the time for the iteration was too short and wished that they would have been able to go deeper into the system. They had implemented the method in the iteration, looked at the two worlds and could see the large task in creating the systems of game play, and style and mechanics between the worlds – the parasite and the carrier. They liked the method and they thought it was a good system. They saw how one could control the information through the method. They expressed that they were happy that they now had a tool that they could use for future work.

## 3.7 Group 4

### 3.7.1 The idea in brief

The title was " Unknown destination" and the premise was about an innocent drug addict that gets accused of murder and has to avoid the police and flee the country. The goal for the user was to feel exposed to a drug addiction and avoid the police and finally leave the country. The game was set in a grey, rainy and dismal city. The character had a first person perspective and the player should feel and understand other's (npc – non playing characters) facial expressions what they thought about the character's appearance as a drug addict and outcast.

The group presented the game play by describing a scene from a hospital where the player tried to find drugs. It was night with not many people around, and the player needed to find keys or break into rooms that were locked. One had to sneak, keeping a low profile to not risk the staff alarming the police. The effects of the drug, withdrawl, were well described, how it obstructed the player. They described the stress, the blurred vision, bad hearing, and difficulty moving. When the player got the drugs, the alarm at the hospital was activated and the player needed to escape. The player needed to balance the drug addiction, and getting drugs to function, as the withdrawl caused problems. Another rule and condition for the game play was if one tried to oppose the police force it would increase in number so it was better to flee and try to reach the airport and leave the country.

## 3.7.2 Simulating the idea with the method

→ Premise/Syuzhet – Innocent drug addict is accused for murder
← Goal/Fabula – Exposed to a drug addiction and avoid getting caught by police.

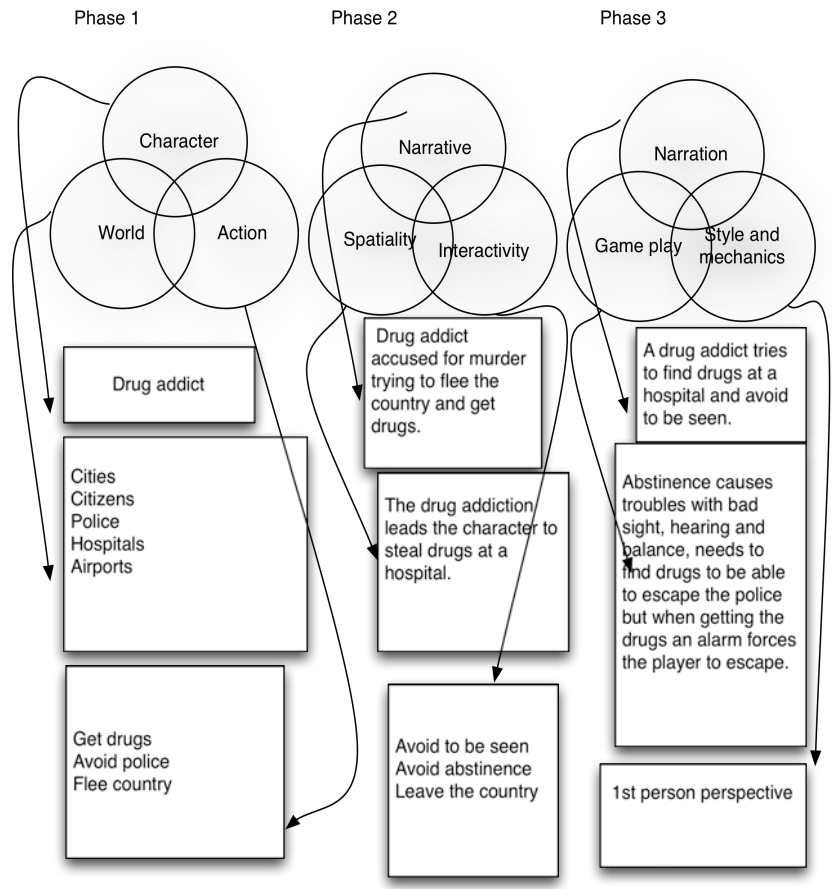

**Figure 21:** Group 4 and the first iteration of their idea.

### Phase 1 The diegetic world

The rules, relations and logic in the diegetic world was set in a city where a drug addict, innocent accused for murder, tried to escape at the same time the character needed drugs. It was a world with hospitals, drugs, hostile habitants, police and airports. The diegetic world seemed balanced and created a clear picture of the space. There was logic in the relation between the character's addiction and how the habitants looked upon the person. When looking at the relation between the police, the murdered, and the character, the question was who was murdered and the cause to leave the country? Simply expressed, what was the logic behind the caused the accuse?

### Phase 2 Interplay between narrative, interactivity, and spatiality

The group had narrowed the space by showing how the causes-and-effects in the diegetic world looked at the hospital when the character needed drugs. The elements of information were linked to each other and created conditions and consequences so the player had no chance to rest as he was "chased" by an inner addiction, the police, and the staff at the hospital. As the goal was to feel chased and exposed, as a player this worked well when it came to the addiction. What was vague was the information about the murder and the police force and with what intensity they chased the character that forced him to flee the country. Another question occurred in the spatiality about the character and how the character moved: Did the character have a driver-licence, or friends, where the character could hide, etc?

*Phase 3 Narration, game play, and style and mechanics*

The group presented a developed game play with a well balanced cause-and-effect that created conditions and consequences such as being chased, trying to escape, needing drugs, risking being found, not fighting the police as that would increase and try heading to the airport. They had developed the style and mechanics when being effected by the addiction that was well related to the game play and how it effected the interaction via a time limit, blurred vision, and get the drugs in time before being too affected which would risk to get caught. One question remained though: whether the group wanted to give the diegetic world a deeper background to the motivational and narrative schemata and retrieve more information for the game play, and style and mechanics?

### 4.7.3 Results of simulation and iteration

At the second meeting, after the iteration, we discussed the further design. The group had a well-balanced game play with narrative templates (the city, police, addiction, etc.). By having a more detailed background they could increase the motive for the player to identify himself or herself with the drug addiction, while also knowing more about whom the character could have murdered to understand the dimension of the police forces. Knowing more about the character would also reveal skills, etc. During the discussion the group said how they had been working with a conspiracy theory in the game and thought about the character's background.

The group handed in questionnaires instead of a new design document and therefore the second iteration has no graphical presentation. In the questionnaire they reflected over their design process, the development of the idea, and the use of the method. From the beginning they wanted to make a game that expressed how life could be for a drug addict and to develop the motivating mechanisms that expressed the situation of being effected by many side effects. They felt that the character was "flat" and needed a better background to become more believable to motivate the user. At the second iteration they found out that two of them had different ideas about how linear the access to the information should be for the player. The method had helped them to structure the information and not forget anything and they felt they could work with the idea on a depth and on a more detailed level. This enabled them to get rid of cut scenes and telling the story. Instead they could implement the narrative with the game play and let the story be told through the interaction. What they would have liked to do, but the lack of time prevented them from, was to elaborate the conspiracy and the game play to not make the player feel it escaped on pure luck and not loose faith in the characterisation.

## 3.8 Group 5

### 3.8.1 The idea in brief

The premise was that the group wanted to explore the adventure game and the firmed structure of a hero's journey based on Joseph Campbell's deconstruction [10]. The goal for the player was to experience a characteristic adventure game, which was to success, and reach a higher self.

The game was about a little boy, a son of a glassblower, who looses everything after an attack on the village by unknown raiders. When seeking revenge the boy almost dies and is found by an old lady that becomes the boy's mentor. The old lady turns out to be a witch that helps the boy to find his magic powers the first mission was to find his grandfather's book about magic. The old witch trains the boy and he moves around the world to find clues about who attacked the village. His power grows and he finds a glass-sword. In the end the traces leads back to the old witch that saved him and became his mentor, who turns out to be the one behind the attack and the one the boy finally has to defeat in a final battle.

The group described the balance between narrative, game play and game mechanics by discussing the beginning of the first act, "Call to adventure", "Refusal" and "Meeting the mentor" and how the boy goes out in the forest to pick up wood and how this was a chance for the player to get familiar with the basic controls and game mechanics. When coming back after gathering wood, the boy sees the village being burned down. The game had a third person perspective. All information should be implemented in the interaction within the diegesis and not as cut scenes. The dialogue should be voice acted and getting new skills represented the protagonist's (the boy) progress.

### 3.8.2 Simulating the idea with the method

→ Premise/Syuzhet – Make a classical adventure game
← Goal/Fabula – Reach a higher self

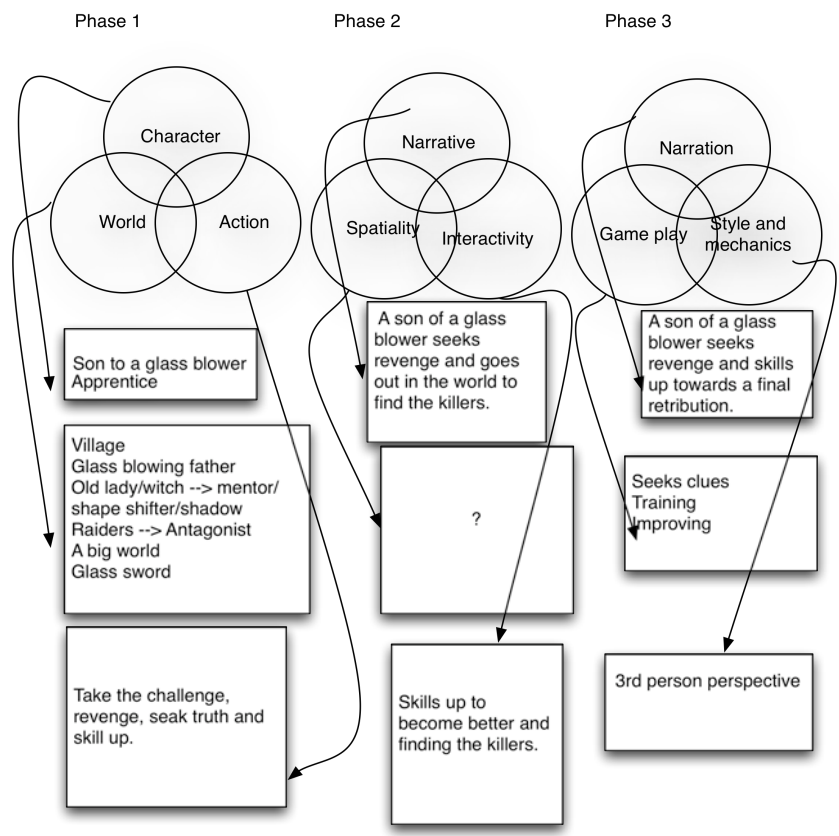

**Figure 22:** Group 5 and the first iteration of their idea.

*Phase 1 The diegetic world*

The group developed the rules, relations and logic to the diegetic world by choosing the action to take place in a world of magic, along with a character that was the son of a glassblower. The boy's background as an apprentice to a glass blower was represented in the world by the glass sword as one of the items the boy needed. The group wanted to explore the strong structure of an adventure game. They created commonly occurring characters that created a drive-to-goal pattern through having a protagonist see his village and family destroyed and after that searching for the guilty and retribution. The main character finds a mentor, an old lady, which later on turns out to be the antagonist (shape shifter), and the one the boy finally needs to defeat. This move made the goal even harder as the boy needed to become better than his mentor to win and reach a higher self.

*Phase 2 Interplay between narrative, interactivity and spatiality*

When moving over the information from Phase 1 to Phase 2 to find out where, how and why the action took place, it showed that the world needed further consistency developed. When looking at the causes-and-effects for the user's spatiality and interactivity a question aroused – who did the character meet except for the mentor? This called for a map of the world to be made telling where people lived, etc. This development was not only to expand the world description but, as the group had chosen to hide information about the mentor (that turns out to be the antagonist), the group needed to mislead the user by creating a "disguised" antagonists if they wanted the player to be unaware about the twist. By elaborating the world the magical powers would be detected by asking how does the player should fight the enemies, which could be done by knowing whom they were. Glass blowing was a main skill for the boy and this was something the group had thought of by making a glass sword. The "glass abilities" needed to be developed by knowing everything about which enemy (the causality) and where to fight. The work was very much about developing the characters and relations between them and the world to see what action came out of it.

*Phase 3 Narration, game play, and style and mechanics*

The group had clear ideas about how the narrative should be presented within the digesis by not using cut-scenes or text. Instead the narrative system should be presented through dialogues in the game and by voice acting. The game should be in third person perspective.

### 3.8.3 Results of simulation and iteration

After the iteration the group kept the premise and goals and stayed with the idea to make an adventure game and continue the exploration of the structure. From the first stage of idea generation, the group had developed the diegetic world and brought more information to the world and the characters.

→ Premise/Syuzhet – Make a classical adventure game
← Goal/Fabula – Reach a higher self.

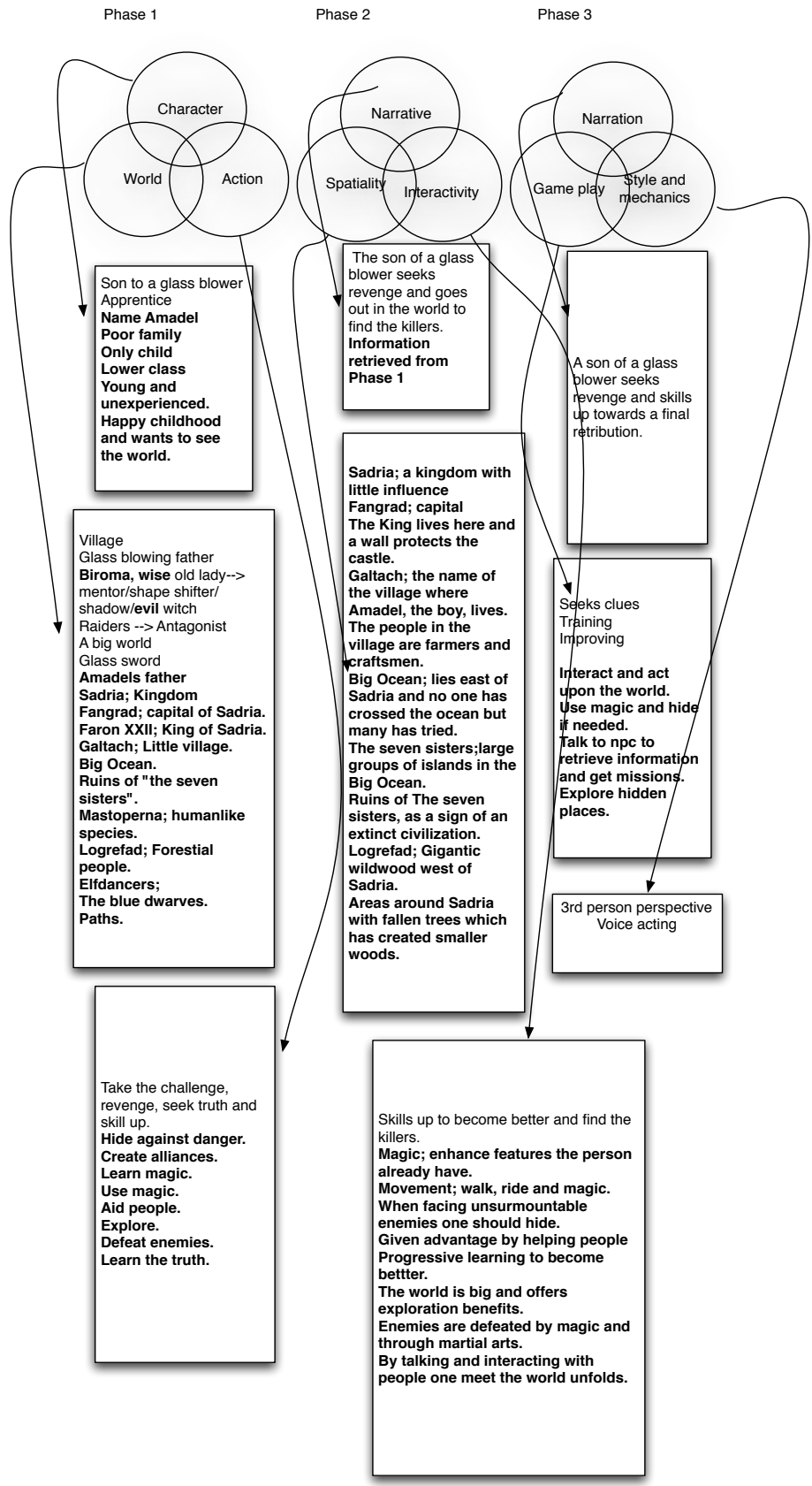

**Figure 23:** Group 5 and the second iteration of their idea using the method.

This time the group had developed the diegetic world in Phase 1, while logic and relations were developed in Phase 2. They created more inhabitants and built a kingdom called Sadria with a powerless king, Fadron XXII. The elaboration and movement of information between Phase 1 and Phase 2 gave a good idea about the protagonist's world, the boy, who comes from a poor village that fed on handcraft and farming. One could see how the group started to build challenges in the world such as the "Big Ocean" no one had ever been able to cross it and it had a group of islands, "The Seven Sisters". Through the information from Phase 1 and 2 the group could present a system for game play for how to interact and act upon the world as hiding and helping people in the world as a part of the game play. But as the antagonist was not fully elaborated it was hard to tell what kind of magic to be used or how the threat looked like that one had to hide from. The group simply laid a good base for further development of the world and the adventure if more time was given to the development. The only thing they had not done that could have improved was to define the motive, the cause-and-effect, for why the witch acted as she did.

Two weeks was too little time for the group to develop an adventure game. They thought the method took too long time to understand in the beginning and it would need to be simplified and more distinct. They referred to the internal relations between the elements in the phases as being a bit unclear. Most of them found third phase hard to manage and how to organise narration, game play, and the style and mechanics. When discussing the development of the idea and the use of the method – after figuring out how it worked – the group found the system good for finding out things about the world by how it arranged the information in Phase 1 and 2. They thought these phases helped them to frame the world and made it easier to communicate, in the group, as well as how to connect the information in the world such as "then one can connect that and that to do that and that". The group suggested the "style mechanics" to be called "style and mechanics" (which was implemented as a term for the original definition *style mechanics*).

## 3.9 Group 6

### 3.9.1 The idea in brief

This group was formed during the first lesson (one week after the others) and began creating a game idea about an ugly male troll in a medieval fantasy world. The premise was a troll that needed to find potency to secure the continuation of the troll race. The game should be humoristic and a satirical. The pictures showed an idea of how the troll looked and they were taken from the films "Shrek" and "Harry Potter" showing a swamp troll and a house-elf.

The troll should start the adventure in a cave and talk to a shaman to get the first quest and the goal was to find the Orb of Potency. On the way the troll should encounter the Dark cave, the Evil castle, the Happy village, the Hermit on the Mountain, the Fallen Star, reach the Moon and explore the Land of Nymphs.

The group wanted to keep game play and style simple with puzzles and labyrinths. They wanted to find a convenient tool as they believed their main teacher would ask them to make a digital prototype of the idea afterwards. If that was the case they thought of the game-making tool "Game Maker".

### 3.9.2 Simulating the idea with the method

→ Premise/Syuzhet – A humoristic and simple fantasy world that one can make a digital prototype of.
← Goal/Fabula – Find potency to secure future of the troll race.

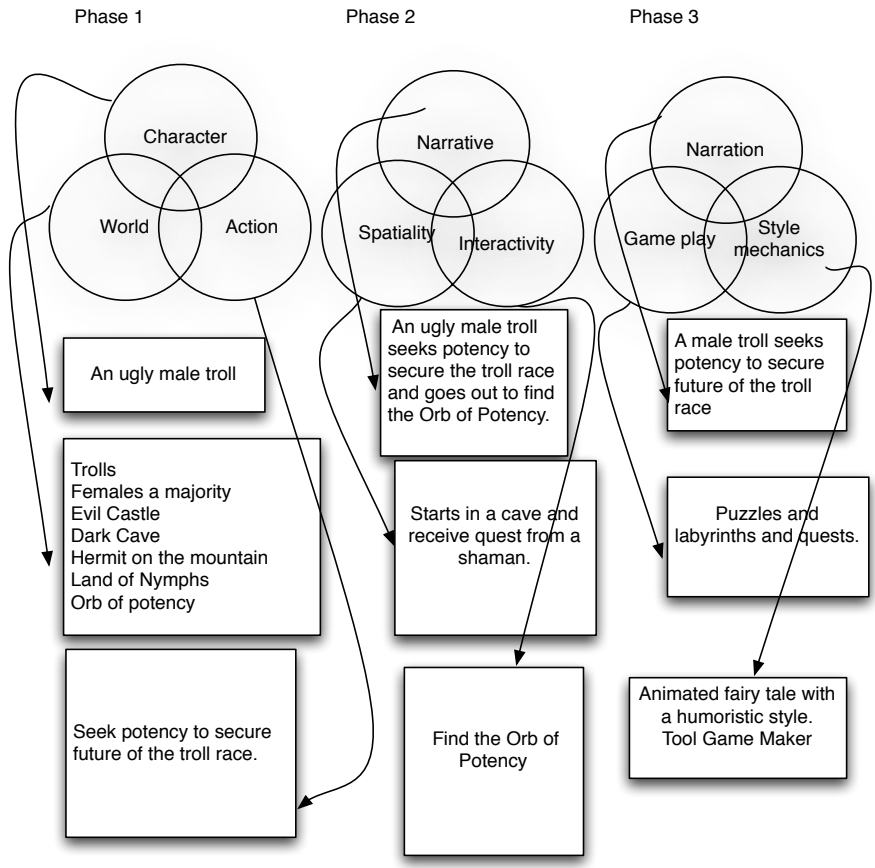

**Figure 24:** Group 6 and the first iteration of their idea.

*Phase 1 The diegetic world*

The rules, relations, and logic in the diegetic world were set in a humoristic fantasy world with an ugly troll that was the only male who lost his potency and needed to go out in the world to find the Orb of Potency to secure the future of the troll race. By making the troll ugly and the only male in a troll tribe, a problem was created that needed to be solved – a clear cause-and-effect and goal. In the world there were females expecting the troll to solve the problem and it forced the troll to go out to the world that was deliberately filled with templates for what can be found in a fantasy world such as elves, trolls, caves, castles, good and evil, etc. The element's construction motivated the player to take action to solve the lack of potency for the troll.

*Phase 2 Interplay between narrative, interactivity and spatiality*

When moving over the information from Phase 1 to Phase 2 the group presented how the troll should go to a shaman in a cave that gave the troll (player) the first quest to find the Orb of Potency. That provided an idea about the spatiality but they needed more time to see how the world and the inhabitants were connected to the troll and what kind of interactivity this created. What one could see was the start of a quest system that could be developed by elaborating the world in Phase 1 by asking where did he go, who lived where, who had the solution to the problem, where were the obstacles to reach the goal of finding the Orb of Potency, etc.

*Phase 3 Narration, game play, and style and mechanics*

The group thought they would have to make a digital prototype of the game so they had already chosen the tool, "Game Maker", to enable the creation of the game. They wanted to keep it simple with labyrinths and puzzles. They also added a quest system to the game play. The problem that occurred for the group with further development was that the simplicity the group wanted for digital prototyping contradicted the fantasy world, explorations and quests. The group preferred to use easy puzzles and labyrinths with quests systems. The group had to decide if they wanted to go back to the premise and think through whether they wanted to make a simple labyrinth and puzzle game so they could make a digital prototype, or develop their humoristic fantasy world. The group could also try to make it simple and try to balance the features of an adventure game. Simply expressed, it was up to the group to decide what they wanted to design for the premise.

### 3.9.3 Results of simulation and iteration

After one week the group returned. Their work had been to decide what kind of premise they wanted to have – make a game that they could easily make a digital prototype of or if they wanted to developed a fantasy game or something else. If they stayed with a premise of making an easy digital prototype, it would also make the narrative style system superior to syuzhet (see section 2.3.2) and they would have to manage that balance. When the group came back they had chosen the premise to make a humoristic fantasy world and left behind the idea of making an easy game to solve a digital prototyping. The goal (fabula) was to secure the troll race but they had change the male troll to a female and instead of finding potency it became a search for fertility.

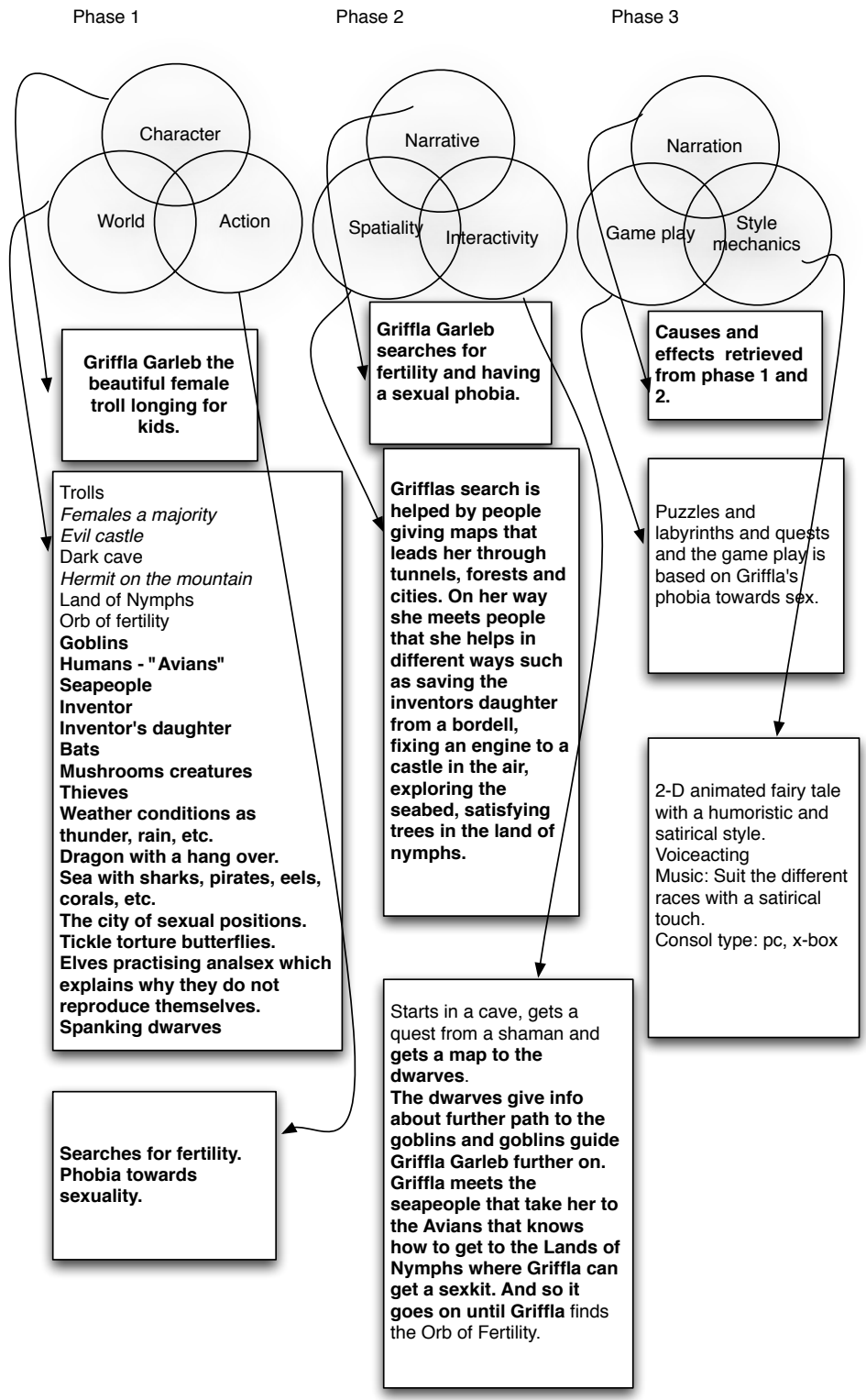

**Figure 25:** Group 6 and the second iteration of their idea using the method.

The group had one week to develop the idea as their group was formed later. The major changes they had done were to make the ugly male troll into a beautiful female troll longing for children who had to find the Orb of Fertility. The new problem, or cause-and-effect, was now that the female troll had a phobia against sexuality. Why the group made this change was that they wanted, as one wrote in the questionnaire, to shock and make a fun game with satire and humour. They thought a male was not as fun as to have a prudish female troll. What they did was to add a phobia to the troll character to use in the game

play – putting her into all events that were connected to a phobia against sexuality in her goal to find fertility. In the former idea they had used the male's ugly look to create a cause-and-effect to the goal but this time it was instead a phobia. The group gathered information to the world connected to the goal for the character and tried to map a system for the spatiality and interactivity. The time was too short to elaborate an idea for how the labyrinths, quests and puzzles would look like but they made a base for it to be created. The group also mentioned that in the questionnaire that, if they had more time, they would have solved that and all other interaction with the npc (non playing characters).

The pressure of time made them move quickly through the method and work consciously with the phobia as a base. The method's Phase 1 and 2 helped them, they said, in the fast process and prevented them from not "standing and stamping on the same site" as the method forced them to move forward. The group found that very pleasing as the progress made them solve the problem somewhere else in the method. Phase 3, they saw, as a distribution of information and did not offer the same working flow as Phase 1 and 2, they thought.

# 4 Evaluation of the results

The narrative bridging method was aimed supporting creation, organisation, control and generation of information for interactive media represented here by digital games. The method was adapted to support new forms, as well as structured and genre-based forms, and enable simulations of an idea.

In this study, the method was applied to digital games and tested by game design within a course in rapid prototyping where the students tried different methods. In this chapter, the evaluations and results will be presented through comparing the iterations, as well as the gathered opinions from discussions and questionnaires to make a whole. The aim is to see if the method achieved the desired outcome and by the results improving and optimising it.

Some of the optimisations have already been done as the qualitative study were preceded by a seminar, a lecture, having discussions, which provided an immediate response from the subjects. One of them was to rename *style mechanics* to *style and mechanics*. The optimisations are made in some graphics showing causality (see Figure 8), and the pointers showing the working practise in the designer's and the receiver's construction (see Figures 3 and 4). Other immediate responses during the seminar when simulating the ideas were "that the practise recalled scriptwriter's and designers work". The comment made me realise that the method mimics the scriptwriter's practise when handling the information to reach logic, relations and rules in a diegetic world. One student asked or asserted that it seemed to be better to elaborate the world and the character first, before the action. This could not be confirmed as the study and the work developing the method had not ranked the systems and their elements. But it said something about the viability of both the syuzhet and the style as systems [7]. If one looks at the game ideas, all groups moved back to develop the diegetic world to create stronger motives and restricted their narrated objects and attributes which formed a stronger game play. A discussion about which of the systems was superior would only lead to the discussion as e.g. the one between game rules versus narration and is not what this work aims for. What it does show is the relation between the cognitive mechanism and the goal driven human. It tells how the narrative practise can manipulate the cognitive patterns to create meaning and how the construction of the method supports this process.

In the questionnaires, about the subjects' knowledge about narration, they looked upon the *disadvantages* from the perspective of narration as a story, something told and linear. They thought narration was bad when it required focus, was forced upon the player, by saying things the player did not agree upon, or interfered with game play. If badly written, one said, the player looses interest and the experience gets static or causes confusion. Some said it took time away from the development. When the subjects answered what the *advantages* were with the narrative, the subjects saw how it could create emotions and empathy, understanding, interest, enhance game play, make it more believable, etc. To substitute the answers they all thought that if the narrative was well-handled it was a strong tool for the creator in game development. The few that were most negative towards the narrative were also the ones with a strong opinion that the narrative represented a story that was told and was linear.

## 4.1 Overview the setting and the method

In general the ideas that did not elaborate the diegetic world experienced problems establishing the spatiality. This effected the interactivity as one did not have walkthroughs or maps and did not provide motivation. The causes-and-effects needed to be elaborated to target the interactivity towards "what" and "where" to motivate for interaction. The groups using templates as an already familiar infrastructure as e.g. a "historical drama in a castle", "city", a "body" (group 1, 3, 4) got a rapid access to the spatiality and the interactivity. What made them move back to model the diegetic world was that the character needed to be developed so the player could see what conditions and consequences the diegetic world produced. In the idea with the parasite, the movement back to the diegetic world was to improve an idea, to go beyond the templates towards an original idea with greater complexity. Another discovery was if the character in the diegetic world (Groups 1, 2, 4) was not developed, one also faced problems setting the interactivity and the spatiality. Even if there is a learning system for the user at the start of the game, the designer needs to know to be able to cue, channel, and manipulate the objects and attributes in a way so the user can learn about their position and world.

All groups had limited time and felt it was too short for handling the complexity of the systems that appeared. At the first oral presentation the ideas seemed to be thought through. When applying the method to an idea, the inconsistency and contradictions were revealed and when iterating a system, its elements expanded. Many groups chose to explore a combination of an embedded- and emergent narrative system. Many chosen to break down the idea and look at a specific event in the game to see how the interplay between narration, spatiality, and interactivity revealed the conditions and consequences for an activity that formed game play. It was also in this final stage the course ended, causing some frustrations.

If the time was an element of disturbance another annoyance occurred when the systems were revealed through the method. The frustration was that they lacked formats (story-board, flow-charts, manuscripts, etc.) to express the systems. The subjects also wanted to see more examples simulated through the method, which would demand a longer course. It was the causality between the systems and their elements that they found problematic. The problem was most significant in the third phase where the narration, game play, and style and mechanics should be organised. These who had problems and were uncomfortable with the third phase were the ones with a strong embedded narrative system. The one that had an emergent narrative did not encounter the same problem. The reason for the different experiences could be seen in sublimate structure the embedded narrative implies which made them not reach the third phase. Their focus had been on the first and second phase, to establish the diegetic world and its interplay. The ones that had no problems with the third phase were the ones solving the establishing of the diegetic world by using a template of a "city" or "body", a fact which provided them with a rapid access to set game play. They were groups with a strong emergent narrative set at the beginning.

The most helpful phases, according to the subjects, were the first and the second. The subjects had no problems working with causality when using the provided working guide e.g. "Why", and "Where" and "How", to detect the paths in the world and the characters attributes to the world and the interaction.

Seen from the perspective of what the students actually did complete the gains were significant. To have a method that reveals a system and its elements is good. The method turned out to help rapid prototyping by detecting missing information and relations through the first and second phase. The method provided an overall view of the game and allowed the whole group to follow the process. It gave the group a vocabulary "Let's go back to Phase one and set the world" or "Let's see what happens if we look at the spatiality in Phase two". One group thought it was positive that the method did not let them stand and stamp at the same site and made them solve the problems by moving them back and forth through the phases. Another group pointed out that the method helped them to avoid "cut scenes" and instead move the narrative into the game with the help of the method.

Finally when holding the lecture, visualising the phases on a whiteboard, when simulating the ideas, enabled the other groups to see the construction of the idea and the inconsistencies. In this way they could follow design decisions and participate in the discussions about what to do. From an instructor's perspective I was not the one pointing out where the contradictions were. It was the method directing and I was simply a guide.

All individual works will now be analysed in detail to see how the method aided or how the method could be improved to future work.

## 4.2 Analysis of the iterations and results of the method

### 4.2.1 Analysis of group 1

The first group was the one making a game about the assassination of Gustav III. It was a system mainly based on an embedded narrative, where the information unfolded on the way through the diegetic world, according to a crime genre. The genre asserts special rules for the construction and that is to create hinder by manipulate the information in a certain way. The group had done that by giving a goal to the player to socialize to retrieve information.

Their presentation gave a very good idea about the diegetic world and the concept. When simulating the idea through the method inconsistency was detected in the player's goal to socialize through the character to retrieve information about the conspiracy to prevent the assassination. When the character (seen from a user's perspective) should move through the spatiality and interact with its environment the method called for information to be developed as one could not tell what kind of relation the character had to e.g. the king and the court. This was important to know to for the modelling of the social interactivity since one needed to use body language (bow, salute, shake hands, etc.). This lead back to the premise and the goal to decide if they wanted to keep the crime genre along with the socializing to retrieve information, or if they wanted to change it. The group decided to stay with the premise and goal and make the character become a musician. This decision opened up the possibility to develop the spatiality and break down the idea to model the dinner event to see how the conditions and consequences were formed.

**Detection** – Inconsistency was detected in the second phase and the element of spatiality, as a condition and consequence of the activity, could not be created. This led the team back to elaborate the character or change premise and goal.

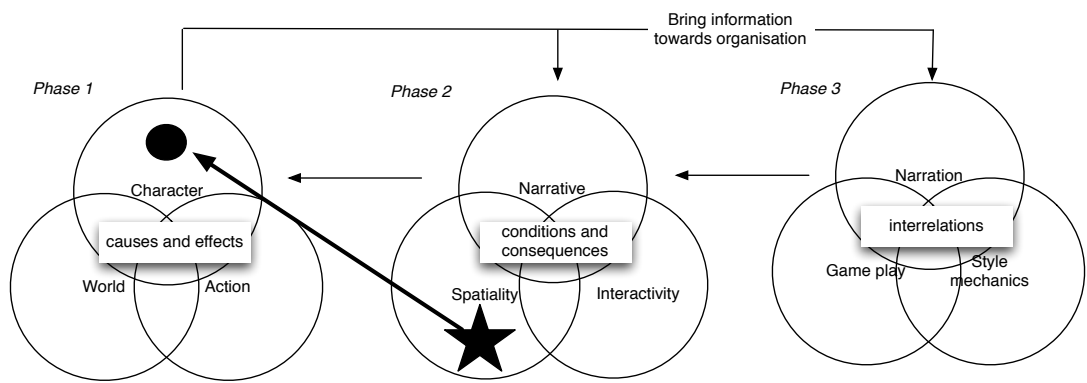

**Figure 26:** Group 1 and detection of inconsistency.

## 4.2.2 Analysis of group 2

This idea was about a girl with a personality disorder feeling disliked and threatened by the world. She expressed herself by writing poems about her dreams, and killed people that "violated" her. This idea was a text and employed an embedded narrative system. This became a challenge for the group to see how to make it into a game, as well as motivate the player to carry out the action. The idea could fit the genre of storytelling but then the information had to be manipulated in a way so the receiver could experience the premise showing a girl's exposure. When simulating and reaching the second phase an inconsistency occurred in the causality. One needed to know how the girl's world looked like, where she went, where the threats occurred, etc. If this was made one could establish the conditions and consequences for game play in the interactivity. If the group wanted to stay with their premise they would need to develop how to express the girl's personality disorder. The simulation led the process back to the first phase and the world. The group decided to divide the diegetic world in order to express the personality disorder. This enabled the group. This allowed the group to investigate how the girl interacted with her mother when asking if the girl had done her homework and how the girl saw her mother as a monster. They lined up what order the interactivity occurred to form a game play between the dreams, the poem and when and where the threat occurred. As the idea was a challenge we discussed how to cue the information to pass on the message to show a child's exposure in the world.

**Detection –** Inconsistency between the character and world were detected in the spatiality, in the second phase, where the causes-and-effects were not developed between the character and the world. This meant that the activity in the second phase could not be elaborated unless one wanted to stay with a condition of killing without any consequences of a condition.

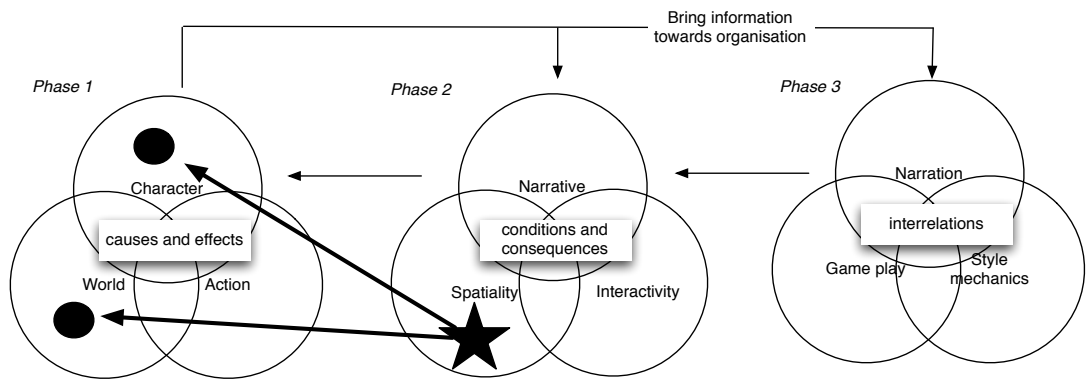

**Figure 27:** Group 2 and the detection of inconsistency.

## 4.2.3 Analysis of group 3

This game was about a parasite that takes over a body for its own survival and prosperity. This idea had strong interactivity and gave a good idea about the game play that had a strong feature of an emergent narrative system. The system was developed with the help of templates, forming an anatomical and the system that would get effected by the parasite. When simulating the idea it went very smoothly through all phases. The information retrieved in the first and second phases could be brought to the third phase where the systematisation and detailed schemata for the game could be scripted and brought closer to computation. But the group also had a "body", a carrier of the parasite, they wanted to effect in the premise and goal. If they wanted in order to keep that idea, the iteration had to go back to first phase and divide the two worlds to work out causality between the parasites world and the world of the person they wanted to effect and who that person was. After the iteration the group presented the "body" as a rude person at the top of his career that they wanted to see fail. By defining the world of the man they changed the spatiality for the parasite as well as interactivity and game play. The targets for the parasite that had been built upon main in the body, as the heart and brain, became instead targets that gave the man compulsive movements that, and put him into awkward situations when attending a board meeting, having a date, or other issues that were important for the carrier. The iteration, by dividing the two worlds, opened up new casualties that effected the spatiality, interactivity and game play to not be based on the templates of a body, instead becoming a more multifaceted scheme and game play by keeping an eye on the carriers activities while simultaneously trying to survive as a parasite inside the body.

**Detection** – Inconsistency was detected in the second phase and the narrative as the premise and goal, to effect not only body but also a person carrying the parasite, was set. Therefore the world needed to be elaborated to create additional information for the conditions and consequences of the activity for the game play.

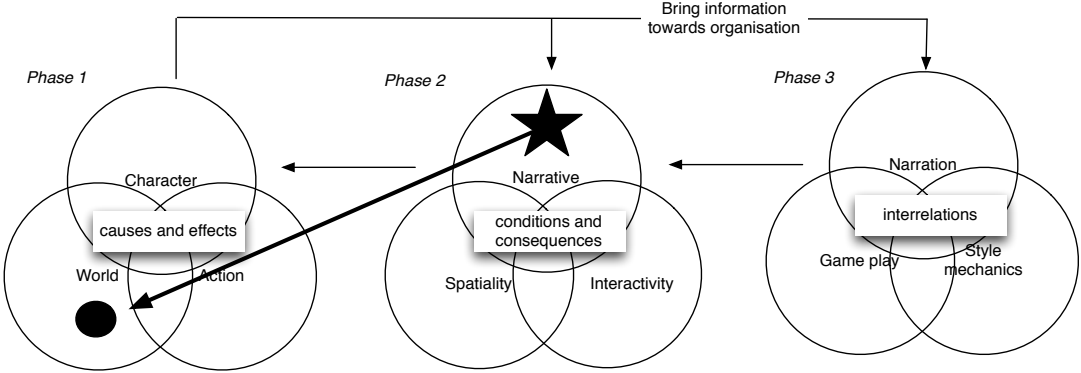

**Figure 28:** Group 3 and the detection of inconsistency.

## 4.2.4 Analysis of group 4

The fourth group had a premise about an innocent drug addict that gets accused of murder who had to avoid the police and flee the country. The goal corresponded to the same. This group had strong interactivity developed by using the template of a city and criminal field for drug addicted. They created strong causalities between the templates and the accusation of the character which created a strong game play for how to move, what made the character move. For instance reach airports that could take the player out of the city at the same time the drug addiction created problematic symptoms. When simulating the idea through the method one could see how the third phase had retrieved lots of information that created a possibility to develop a scheme for game play to be systematized for computation. The question was Who was the person that had been murdered? Did the threat only come from the police and an inner and outer conflict confronting a drug addiction? Even if game play were strong, the logic for why the man was chased would have risked a decreasing of the motivation, as the genre demands an explanation. It would also help the team to know how to systematize the conditions and consequences for the game play if the causality was further developed. The group faced the same choice, as the third group that the interaction and game play could be more developed if the causality between the elements in the diegetic world were elaborated, even more in this game as it belong to the a crime and action genre.

**Detection** - Inconsistency was detected in the second phase and the narrative, as the causality was not elaborated and the logic for being chased would offer additional conditions and consequences to the game play.

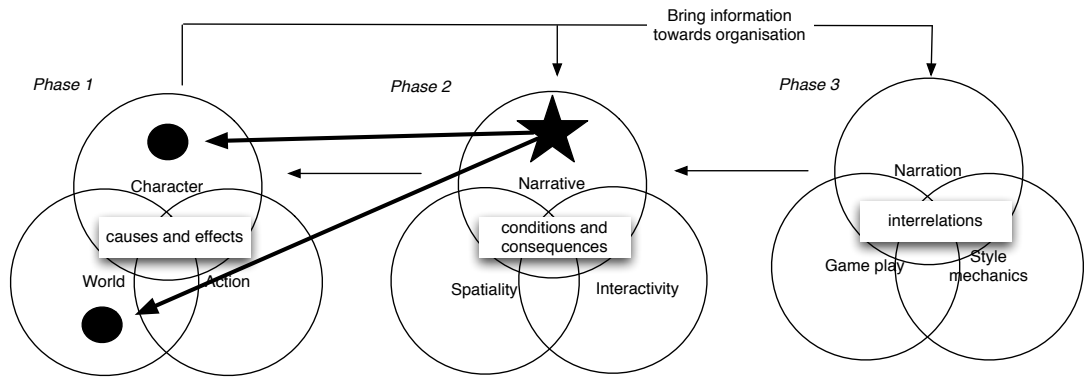

**Figure 29:** Group 4 and the detection of inconsistency.

### 4.2.5 Analysis of group 5

This group wanted to explore an embedded narrative system by making an adventure game. This meant that they had to keep a strong structure that was genre based. They used the "hero's journey" where the goal for the player is to reach a higher self. The idea was about a glassblower's son and when simulating the idea through the method the first phase was due its genre more developed than the other parts. The group accepted a huge task developing such a large diegetic world. On the other hand, the structure for it was already set. Further work would be, for the second iteration, to structure the world map. This would be helped by using the spatiality which would reveal the interactivity, e.g. what spells the boy would use, where did he meet the different challenges, and who inhabited the world and where were the hinders and higher forces that he needed to defeat. This work would also be helped by developing causes-and-effects for e.g. the mentor's motive. The group was well aware about the work that was revealed by the method but they took the challenge and started to make further developments with the world, revealed for the second phase and got quiet far with their iteration. With help from spatiality they developed the world-map further revealing the interactivity and a possible game play.

**Detection** – The inconsistency was detected in the second phase and the spatiality that pointed back towards the world in the first phase to elaborate the diegetic world.

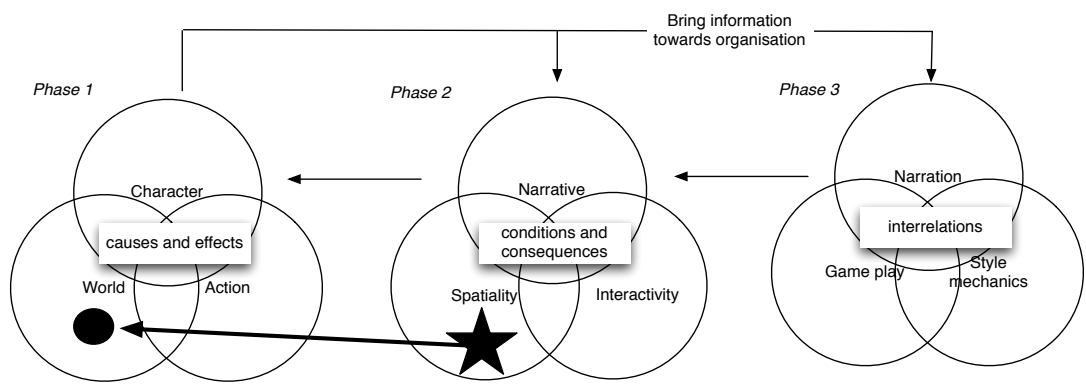

**Figure 30:** Group 5 and the detection of inconsistency.

### 4.2.6 Analysis of group 6

This group was formed later than the others and that was interesting to see as their working process shown without any earlier parts in the process, where the premise and goal went through several changes. Their first idea was to make a humoristic troll fantasy world. They wanted to keep it simple with puzzles and labyrinths as they thought their main teacher would give them the task to make a digital prototype afterwards. The simulation of the idea turned into a problem as the premise and goal contradicted each by the easy prototype versus letting the player experience a fantasy world complicated the process. When the group came back from their second iteration the premise was changed to fully explore a troll fantasy world, eliminating the idea of making an easy digital prototype. They made a gender shift to make a male troll into a female and they created a phobia towards sexuality. This laid a foundation for the spatiality, the interactivity, and further the game play. The phobia made the group create causalities by letting the female troll encounter all kind of inhabitants that could create conditions and consequences for her phobia and this formed a pattern for a game play. The group said that they got helped by the method to rapidly organise and develop the idea, despite the short time.

**Detection** – The conflict in the premise and goal lead the group back to balance these two to enable stringency to the whole. An inconsistency was found in the second phase that directed the group back to first phase as the world needed to be developed further.

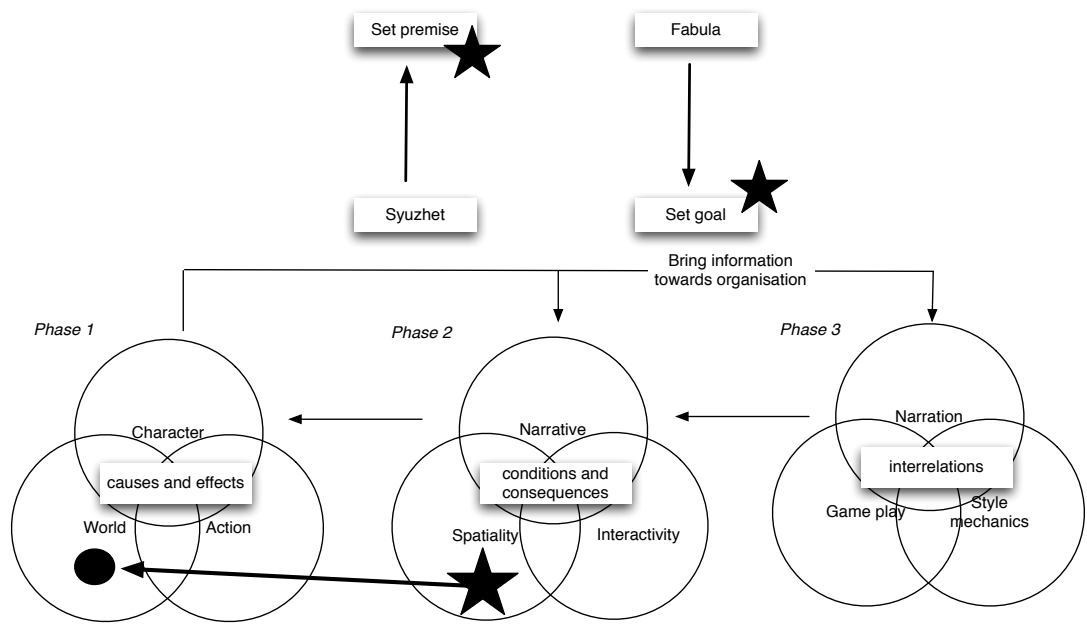

**Figure 31:** Group 6 and the detection of inconsistency.

### 4.3 Summarizing the results

All the inconsistencies detected would probably have been detected sooner or later by the subjects if given more time. But the fact remains that the method gave an immediate response to the idea in a time-constrained design process. The analyses pointed out the main contradictions in a particular iteration. If analyses covered a longer process several problems would have been revealed. Simply expressed, it is the construction of the method and support of an interplay pursuing conditions at endpoint as e.g. the creation of paths and habitants that calls for development of the character and this iteration continues until the causes-and-effects reach logic.

Two groups had developed strong goals for their characters as in the adventure game about a glassblower's son and the female troll (Group 5 and 6). The characterisation of the glassblower's son was treated within the strong structure of a "hero's journey" [10] and was given a drive-to-goal through seeing his village burned down (the so-called "call for adventure"). The group that had a female troll (Group 6) developed a character with a phobia and the drive-to-goal was a search for fertility. The phobia created an engine, setting the causality for the whole game as a search for that created a clear conflict and hinder that inflicted the setting of the causality between all the elements in the phases. It tells something about how the development of character can create a causal machinery to cue and manipulate the information to form a game play. The problem this group faced was a conflict in the premise (digital prototype versus fantasyworld) but was captured by the construction of the method that had the premise as a directory through the whole process. This showed how strong the premise and goal for creating a clear direction towards a cohesive game and it also showed how a second stakeholder could influence a design situation as e.g. market interests, distributers, producers, etc.

Another example where one could see the importance of an elaborated character was in the game about the assassination of Gustav III (Group 1). It was a crime genre where the player by social interaction should find out who was guilty of murder. The method detected that the character was not developed, which showed that there were no conditions and consequences (Phase 2) that could set the game play for socializing. Even if the group wished the player to learn about the character when starting to play it is still important for the designer to know all about its character to model the learning system. When reading the pitch, many ideas seemed to be very thought through. This says something about the pitch as a format and how it can veil design but to run an idea through the method could give the pitch depth. The group that made the assassination game solved their problem by creating a musician. In an extension of the development one could certainly have expanded the variety of characters at the court for the player to choose between. If the group had chosen to not develop a character the game would have become more of an historical documentary where the player would click through the world.

Another example of a character that was presented (Group 2) which had a strong literary characterisation was a girl that went berserk with a knife after perceiving the world as full of threats. If comparing the girl with the glassblower's son and even the troll, thinking in terms of creating drive-to-goal there were no motives presented other than a personality disorder. What the group would need to do was to create the motive by showing the "real" world to mirror the girl's disorder, just as in how the glassblower's son and the phobic troll had their motives created for the player to take action.

None of the groups presented an idea where the player could create its own character with different abilities, an avatar [14]. The game about the assassination of Gustav III invited this as it appealed to a world with many inhabitants and actors with different skills and occupations and they all had a joint concern for their king. Even the game with the parasite had the possibility to create several kinds of parasites with different skills. The other games that did not invite creating several avatars as their causes-and-effects more tightly connected to the character as the drug addicted person accused for murder, the troll that searched for fertility or the glassblower's son that saw his village burn down. On this basis one can say that the more one knits the drive-to-goal to one single character, the trickier it becomes to create an emergent narrative with an avatar(s). It would be possible to create a change between characters for the player as in the example with the glassblower's son if he shared the destiny with a sister or someone else. This also speaks to the definitions of emergent and embedded narrative systems [27, 30] that they need a refining that goes more into depth for what it is to elaborate a narrative system for a game.

In the results there were two groups that had strong interactivity and game play from the beginning (Group 1, 3, 4) where causality was based on templates of a "body" and a "parasite" (Group 3) and a "drug addict" being chased in a "city" (group 4) and the historical drama of Gustav III (Group 1). The templates, the body and the city, gave the team a rapid infrastructure for the diegetic world and immediate access to the interactivity and a game play to be set. The groups had strong game play from the beginning and a pitch that looked complete with a developed style and mechanics as well. The problem was that the idea with the drug addict (Group 4) had a template of a city to model but lacked information about who was killed and caused the pursuit. Even if the player would be kept uninformed about whom he was guilty murdering the creator would be helped to know how to cue the information to settle the search and the chase for game play. By doing this the player can sense the causality and can see the consequences of their choices. In the example with the parasite that also had a strong interactivity and game play (Group 3) they had a template of a body. This template set the game play for what to attack (heart, brain, etc.) and what was threatened (the immune system). By developing the body to become a person, that was not very pleasant, the goal and premise extra layers expanded the complexity of the game play. The group could have stayed with the body but what made the idea stronger, by setting the goal to affect a bad guy, created a deeper logic to the player to go ahead and effect rather than infect the body. Finally the historical drama about the assassination of the king (Group 1) had an already set infrastructure that made the group to finding a game play of a socializing character. Their process where pointed back to the character.

Most of the contradictions were detected in the second phase, and the spatiality. Games represented by the embedded narrative got help from the spatiality to create an idea about the world and the interactivity. The embedded systems were also the ideas that did not reach the third phase due to the short time. The ideas with a strong emergent narrative reached the opposite result and needed to elaborate the narrative. It was harder to detect inconsistency in the ideas with an emergent narrative using templates as they gave a clear idea about spatiality, and interactivity in the first iteration. These ideas could have stayed with the templates of moving around in a city or a body but when taking a closer look the motives were vague. Both ideas reached a deeper logic when modelling the diegetic world. By looking at the method from the emergent and embedded narrative, the method balanced both systems. It also handled a literary text (Group 2) in that it guided the process to convert a literate text to suit the interactive media.

The method helped all ideas to detect inconsistencies in an early stage of the process so they could be taken care of and treated from the aspect of the desired premise and outcome. The solution for how to take care of the detected inconsistency lead the process back to the setting of the premise and the goal to see if they corresponded with the causality modelled through the systems and its elements. As said in the beginning, most of the ideas might have come to the same conclusions that the method pointed out, but the important result is that the method aided and in earlier detection. Without intruding, the method gave a hint where to go, and directed by the premise and goal, the groups could take design decisions. How the method aided the rapid prototyping was by its organisation of the systems and especially with the help of the vehicles aiding the causality, as the syuzhet principles offered. Asking why, where, when and how is a natural way when working with problem-solving and is nothing new. But the division of the systems into three phases, offered an overview and a vocabulary for the group to navigate together through the material. Even by keeping the perspective of the user through the method made the design team to take the character's position and orientated through the phases without distracting themselves by the external activities until reaching the third phase.

The most important vehicle creating the interplay was the causality provided by the syuzhet principles. The three phases provided a tool to organise and control idea. As stated earlier, the method generated information and it pushed forward the creation of information, as the subjects made the generation of the information by themselves. Instead, one could say that the method "called" for the information to be generated and what information should be withdrawn.

# 5 Conclusion

The study yielded several results and first a general conclusion about these results will be presented.

First it should be stressed why the method was created. One reason was to pave the way for visualizing the strength and workability of the narrative in the fields of education and research, as well as providing the game design business with a tool for exploring new expressions. The motive for making a narrative method was the lack of knowledge about narrative as a process within the field of interactive media, and a narrative bridge had to be created that did not diminish the properties of the digital games. It was also an issue to show narration as a process and not a "story" and why it should not be treated as an appendix to the other systems. The idea about narration being a story and something told also influenced the students that looked upon the narrative as something being told. But the interesting fact was that when talking about the advantages with narration they talked in terms of a process and how to apply the narrative in a "right way" and expressed a belief in the narrative's strength to enhance the players' experience.

The results of the method and the test, showed how the narrative process cued, channelled and manipulated objects and attributes. These "bricks of information" were distributed and aligned to suit media and outcome: in this case the digital game. The premise and goal, and the two narrative systems as the syuzhet and style, guided the process between the systems and their elements. One could see very clearly how a conflict in a premise caused problems by extension when a group tried to combine a fantasy game and an easy way to make a digital prototype. When I told the main teacher about the conflict in the premise he admitted that he had that kind of plans originally but changed his mind so as not to influence the process.

The systems were based on the media-specific attributes for the interactive media defined as user, interaction and space. Secondly, the definitions of the media were transformed to represent the narrative by the diegetic world: the character, world and action. It turned out to help the setting of the information and as one group said, it helped them avoid "cut scenes" and "text" and instead render the story material within the game. This shows how the method supports the motto: "*Do not show, involve*", meaning to let the player take part of the narrated world by interacting with it instead of reading about it (see Spatiality as a system, section 2.2.2).

The construction of the method was a result of looking beyond strong culturally structures of narration as it invoked the possibility to work freely with the interactive media to see what could be "found beyond". An example of how structures influenced practise, even if the method offered an open structure, was how the characters were elaborated. In the evaluation, one could see how the use of a strong structure tied motives to one single character. The tighter the drive-to-goal was connected to the single character, the more unlikely it seemed to open up for several characters to share the same motive and for the player to choose between different types of character. By sharing a concern or a conflict with a bigger group in the diegetic world invited for several avatars to be created. These kind of phenomenon allow for the discussions and design to move beyond a simplification of narration of being linear or a structure. Instead one can explore how to solve these junctions by for example creating games where it is possible to experience an event not only from the protagonist point of view. One could for example start experimenting with several perspectives in a game and first play as a protagonist, swap to the antagonist that becomes the protagonist by telling its version of an event. This might look like a very complex system to create but if consciously organising the scripting it should not be harder than anything else. It also invites exploration of more complex social events as a personality disorder, love, etc. No groups developed a multifaceted character or an avatar but the time was also short and one cannot tell what would here happened if the groups given more time. Neither could one tell what would happen if the method had been presented from the beginning or if the instructions had been different. But it points out that there need to be more added to the discourse to go beyond the strong structures to expand those patterns of thoughts.

Finally, the following came out of the use of the method, the study and the responses.

### – *How the method aided an overview and vocabulary*

The intention with the method was to aid the viewing and channelling of the information. By dividing the systems and the elements the method turned out to aid the team to *overview* the design and the complexity of the media. The three phases gave the team a *vocabulary* so they could follow each other's reasoning and track inconsistencies and find solutions. By using the graphics of the method the idea could be drawn and lined up on a wyteboard and *visualised for others*. This can be useful when several stakeholders want to forward their opinions or having questions about the idea, process or the production.

### – *How the method aided progress by detection*

The divided phases and their characteristics also helped the team not to get stuck spending time on a specific problem. Instead the method *made the team progress*, move on, to encounter the solution to a problem elsewhere. The method's feature was helpful where the groups needed to see the relevance of information to create causality and logic to the intentional premise and goal (Phase 1 and 2). The method's strength was the *detection of inconsistency* in the logic. The method *did not force* any team to solve their ideas according to what was detected. Instead the method offered the team to solve the detected inconsistency by either revaluating their premise or goal, or both, before deciding to stay with the idea or change the values before going back and iterate the new information through the phases.

### – *How the method called for information*

Instead of saying that the method *generated* information, the method *"called" for information* to be created and "served" independently all kind of concepts, genres or ideas. This was aided via the causality, setting the logic, relations and rules in the world that caused the "need". The ideas where templates were used, the groups got immediate access to the complexity of the elements of interactivity and game play (Groups 1, 3 and 4). These groups, on the other hand, needed to go back to add information in the diegetic world to model the causality. The groups that had the opposite, a strong structure and created a new infrastructure for a diegetic world, did not fully reach the development of game play and its style and mechanics (Groups 5 and 6). They were all directed back to the diegetic world. The groups using templates as a "body" and a "city" and a "historical environment" could have stayed with the templates but by taking the idea further, beyond the templates, they reached a new layer of motivation and ingeniousness. Whatever one use to do, creating a new infrastructure or using templates, the time saved depends on the premise and goal. One could say that the use of templates give a fast access towards systematization but what would be lost might be a qualitative complexity of an experience. And the time to create a more satisfying result would be negligible.

*– How the method supported a natural process*

Stefan Grunvogel [15] criticized formal models for forcing designers into standardized workflows and not integrating them in a natural language, as they need to be learned. When watching the work with the method it turned out to have a feature of directing the subjects towards different elements that needed to be taken care of. According to the responses during the seminars and the discussions, the subjects found the reasoning, when using the method, mimicking their own practise when thinking about the design or as one asked, "if I could confirm if it reminded about a scriptwriter's practise?" This leads the answer back to the making of Silent Hill [25] (see section 2.3.2) where the two narrative systems of syuzhet and style were used, even if unconsciously. It says something about the strength behind the narrative systems of syuzhet and style. The cognitive based reasoning behind the syuzhet principles that generated questions like why, what, where and how, to reach causality, could be the explanation to why the adaptation of the method supported the practise. Nor did it interfere with the creative design situation when modelling the contents.

*- How the method treated systems equally*

Another aspect about how the construction of the method mimicking the creative thinking brings forth another question that touches upon the importance of the *narrative as a process* and the urge to treat it equally to the other systems as the game play and the technology. The construction of the method had an order in between the systems, influenced by the narrative theory where the syuzhet, as a system, was superior to the style. When processing the information in the elements of the second phase all processes moved back to the setting of the diegetic world in the first phase. This mechanism was triggered when not being able to answer "what" someone met and "where", in the spatiality. What the method did in this stage was to "call" for information to be created in the diegetic world. This directed the subjects to return to the first phase setting the causality before trying it again in the second phase, drawing the walkthroughs and maps of the spatiality, and the conditions and consequences for the interactivity. To not push this further into the unbalanced discussions about whether if games are narrative or not, which is irrelevant for this work, the result can show why narratives shall not be treated as an appendix to the other systems within interactive media.

*– Time aspects*

There was a time aspect on the whole study and process. The students expressed a frustration and wanted more examples of narrative construction on a semiotic and interpretive level but that would have required more time and practise. They also found the causality troublesome on a theoretic level even if they used it in their practise, which made me rethink how to express and adjust the phenomenon and the graphics. The other frustration was about the bigger system the method revealed when reaching a certain point in the modelling where the game systems were revealed (the third phase where the material was organised towards scripting and computing). But the short time given to the project also showed the viability of the method supporting a concept to be elaborated to an idea that reached the third phase where the subjects could have started "drawing" the systems of a fully workable game. It said something about the method's capacities supporting the control of the contents, detections of inconsistency, "call" for information that was missing, progress, reviewing, and also giving the team a vocabulary to enable them expressing their progress and development to other participants.

*– How time, several stakeholders influenced premise*

The time aspect can also be studied from the perspective how the ideas where framed. One group thought they would get the task, after I had left, to make a digital prototype of the idea. They had a problem in their premise as they wanted to make a fun fantasy world, but they also wanted to make an easy game and enable the digitalisation, at the same time they wanted to complete the task given from me. This stresses how important it is in a design to be aware of how it affects the premise and the rest of the idea. Here the method can help by showing the effects of having a vocabulary and possibilities to overview for several participants.

*– Effects by the use of templates and structures*

The ideas built on templates or strong structures gave a feeling of being well thought through at the first sight of the concept. The pitch was also a cagily format and together with pictures it gave a feeling of an idea being completed. It was only when simulating the idea through the method that inconsistency was detected. In the game ideas that used templates, it was even harder to detect the inconsistency. They had an immediate access to a whole diegetic infrastructure by using a "city", a "body" and a "historical place". This approach enabled the groups to form the game play and when presenting it, the idea gave an even stronger feeling of being ready for final systematization, scripting and computing. This was a very instructive discovery as it says something about why digital games are based on templates and strong structures. To design a digital game is a compound and time-consuming practise as the media is complex in its form and if one needs to get a rapid overview the templates allow the designer to rapidly see if the idea is solid. Even the use of strong structures frames the work so one can elaborate the rest of the systems. The use of genres, defining the game concept, also helps the team that can say: "We are working on an adventure game" and in that way one can get stakeholders and others to relate and join in.

There was another type of structure used under time pressure and more related to the digital game. Two groups mentioned that they wanted to use labyrinths and puzzles. One was the group that was formed after the others (group 6) and the other group where the ones with a literate text that was converted (group 2). Both groups experienced various forms of stress but the fact that they both mentioned puzzles and labyrinths says something about how easily we all pick up known structures just to deliver something.

*– The lack of standardized formats*

The pitch as a format has been mentioned and that was the only format used within this study. The pitch is a cagily format, as it gives a feeling of an idea, but if one looks at the pitch as it was the tip of the top of a triangle it says something about the need to develop the whole to give the pitch a sense of having a solid foundation. The lack of formats became an issue at the end of the study as the ideas reached a stage of getting access to the bigger systems of systematized narration, game play, and style and mechanics. The method did not provide that but a need and a lack was seen that says something about the education and business needs *standardized formats* as one can see in film business. Some formats within game design are borrowed from the film business, as synopsis, pitch, manuscripts, etc. But here, as well as with the method, these formats should be adapted to the interactive media.

## *– Improvements of the method and future usage*

Even if time pressure and the systems of the emergent and the embedded features made the groups model different phases and faced different problems it can not be ignored that the appearance of the method needs improvements. Causality, and the interrelation between the "circles" of the Venn Diagrams confused some. Even if the targets worked with these interrelating parts of the diagram by asking "who", "where", "when" and "why", some of the subjects did not fully see the connections. A new layout and description is needed of the method to, e.g., base it more on the questions and using other features for the phases. One could also think of removing the third phase that confused the ones that had embedded systems but that would instead confuse those with emergent systems.

It must be pointed out, once again, (see section 1.5, Delimitations) that the subjects were represented by a homogenous group of students. Even if the subjects were skilled, other results could have been produced from detected the method on a real design situation within the game design business, for instance. So to make a conclusion about the method and the target group one can see how it can work within education, with some refinements. A future step would be to bring it to the business of game design to see its viability and see if, and how, it could be implemented.

# 6 Discussion

Jenkins thought game designers were well schooled in computer science and graphic design but needed to retool a basic vocabulary of narrative theory [20]. Jenkins had a score with that but the question is how designers should be retooled and if teaching is an answer?

Teaching narrative, I use to say, is like putting on certain glasses on people that disable all kind of future innocence about what narration does and can do. Therefore I found it important in the study to hand out a questionnaire before hand to see how the subjects looked upon narrative. After finishing the project they were asked if their view on narration had changed. Surprisingly all subjects replied that their view about narration had not changed. It made me go through their first answers and when comparing the answers they simply said: if the narrative was badly used, the experience was bad, and if well applied to the media, it was good. That was so true. The narration was simply a matter of applying it in a good or a bad way. But what was the good way and why were designers occupied by structures and templates?

When iterating and simulating the game ideas all information was already there and the issue was simply to forward or withdraw the information created by the subjects and the method helped them processing this. The student's knowledge was based on how to create digital games and the narrative was based on their knowledge and ideas about the narrative's power and weakness. They knew intuitively about the narrative and seen from a cognitive aspect it made sense as an evolutionary cognitive vehicle for our entire search for meaning in our goal to survive. What the students asked for during the study was more practises, narrative practise, and how to narratively model the pieces of information seen from a process and a semiotic aspect and how to put these together into a whole.

Another important issue that came forth during the study was the use of templates and strong structures, even game specific structures as labyrinths and puzzles. In the evaluation one could see how structures and templates were used to get an immediate control to see the durability of the idea. These ideas where also the once that, at the first sight, gave a strong illusion of being worked through when presented but when taking a closer look with the use of the method the illusions burst. The use of strong templates and structures turned out to aid an immediate access to the whole system of narration and the game play and could be seen as a possible reason for why designers seem to be occupied by structures, template and technical issues as graphic, and programming. Here could the method aid and support the practise by its feature offering a rapid overview, control, detection as a method calling for information to be produced or valued. The method could offer a comfortable way of elaborating ideas towards a more enhanced and motivating experience to prevent a team from rapidly framing the information to see its carrying power and end up with a result that is not worked through in its motivational engines. Or it might become one of the one-sided games seen at the market.

The method still needs to be lectured and refined but as the awareness about how to process the narrative for the digital games grows this can change. The theoretical feature needs to be eased up to make the method and its instructional guide more accessible and user friendly. Improvements have already started by making the feature easier to overview and the interrelating depictions more clear to assure that others can understand it clearly without supervision. These new features need to be tried out to see what works best before established. This discussion also stresses the importance for research within forming a theory of game studies to add a little before the game business when it comes to understand schemes and to explore new expressions and not being too occupied by doing research upon games influenced by the business and strong cultural structures. For we must not forget how it all started at Massachusetts Institute of Technology with a team making a duel of two spaceships in 1962 (26). One recommended step towards a progress would be to start look upon the narrative systems equivalent in importance to the game play and the core mechanics. Hopefully this method can offer a narrative bridging by showing that the narrative not necessarily diminishing the media specific properties of the digital game.

To substitute the needs for future exploration of the narration within game design the practise seems to be needed for how to integrate narration into the digital games and offer tools and formats that aid the process. This is easier said than done, as the business of game design has no time or money for practise or internship. As a member of the management team for an education[1] cooperating with the Swedish game design business, the business want to educate people that can walk straight into the business without any internship. Seen from this perspective I think the practise and the development of tools and methods needs to take place within the academic world and within education. The academic world is generous in its constitution and one shares ideas and work together to progress. Within the education are future game designers and there they can be given sustainable tools, confidence and awareness about narration as a process equal to the other systems. This demands that the research within digital games also starts talking about narration with right terms and start separating narration into areas of structures, process and representation to reach progression and enable an exploration of the future games and other platforms within interactive media. This can even take future games towards new expressions beyond unilateral experiences by fully use the high potential the interactive media offers and even express complex themes such as love.

## 6.1 Future studies

This study only covered a small part of the conceptualisation of an idea but produced several unexplored areas. To substitute a few it would be interesting to follow a whole process from a concept to a final result and produce tools and formats as an expansion to the method. Further, this work provided important results about game design, practise and how to precede in future work to take game design and the media towards new explorations. The hope is that this work will inspire more to do research within the area of the systems and elements presented here as the syuzhet and the style in relation with the game play. One hypothesis that occurred was how the action and interactivity, in phase one and two where empty, if the character, world, or the narrative and spatiality were not providing information to the action or interactivity. Is there a ranking or systems behind these that could give further guidance to game research? There were also results about the design process as the use of structures to get a rapid access to the complexity of the game that would be interesting to look at, as well as the area of genres and concepts as embedded and emergent narrative and how to elaborate these definitions in a broader sense. Additional it would be interesting to go into depth and improve conversions between media as when film is made into games. Even the handling of the space and how the information is distributed when it comes to background stories, dialogues, quests, texts, and other information systems needs a study.

I like to wed future studies to the narration as a process and involve the representation and semiotic perspective, which this thesis and the construction of the method touched upon several times. The semiotic within games is an area that has been defined by David Myers [26] that talks about spaceships being symbols of moving objects and calls the phenomenon "aesthetics of play" where the significations of the symbols gets their values from the game play [30]. There are lots to add to the discourse and I like to bring conquered knowledge to the discourse of the semiotics in games and the interactive media.

---

[1] Futuregames Academy URL: http://www.futuregames.se/

I would like to continue a mapping of the motivational engines to see how the semiotic junctions and nodes create a dynamic when applying narration as a process. In an extension I believe it can provide algorithms to create a more delicate AI for a more dynamic diegesis like e.g. development of non-playing-characters (npc) and their movements when interacting with the player in digital games. Finally my aim is to always go beyond current expressions and next goal is to go deeper into the user's position (characters, avatars, etc.) to map and improve the experience and the motivational vehicles that involve the receiver.